\documentclass{aa}

\usepackage{graphicx}
\usepackage{txfonts}
\usepackage{xcolor}
\usepackage{booktabs}
\usepackage{multirow}
\usepackage{multicol}
\usepackage{siunitx}
\usepackage{array}    
\usepackage{dcolumn}  
\usepackage{ragged2e}
\usepackage{ulem}
\usepackage[colorlinks=true, allcolors=blue]{hyperref}
\usepackage{float}
\usepackage{needspace}

\newcommand{\gaia}{\textit{Gaia}}
\newcommand{\rbirth}{$\langle R_{\rm b} \rangle$}
\newcommand{\rgui}{$\langle R_{\rm g} \rangle$}
\newcommand{\teff}{$T_{\rm eff}$}
\newcommand{\feh}{[Fe/H]}
\newcommand{\age}{$\overline{t}_{\star}$}
\newcommand{\mjup}{$M_J$}
\newcommand{\zmax}{$\langle Z_{\rm max} \rangle$}
\newcommand{\eccentricity}{$\langle e \rangle$}
\newcommand{\amax}{$a_{\rm max}$}
\newcommand{\jr}{$\langle J_r \rangle$}
\newcommand{\jphi}{$\langle J_{\phi} \rangle$}
\newcommand{\jz}{$\langle J_z \rangle$}

\defcitealias{Dantas2025a}{Paper I}
\defcitealias{Dantas2025b}{Paper II}

\begin{document} 

   \title{Probing the origins}

   \subtitle{III. Exoplanet demographics across Galactic birth radii}

   \author{Juan José García-Delgado\inst{\ref{aff:alcala}, \ref{aff:complutense}}
          \and
          M. L. L. Dantas\inst{\ref{aff:puc_ia}}\corrauth{mlldantas@protonmail.com; mlldantas@uc.cl}
          \and
          Isabel Rebollido\inst{\ref{aff:cab}, \ref{aff:esa}}
          \and
          Rodolfo Smiljanic\inst{\ref{aff:camk}}
          }

   \institute{
              Space Research Group, Universidad de Alcalá, 28805 Alcalá de Henares, Spain \label{aff:alcala} 
              \and
              Facultad de Ciencias Físicas, Pl. de Ciencias 1, Universidad Complutense de Madrid, 28040, Madrid, Spain \label{aff:complutense}
              \and
              Instituto de Astrofísica, Pontificia Universidad Católica de Chile, Av. Vicuña Mackenna 4860, Santiago, Chile \label{aff:puc_ia}
              \and
              Centro de Astrobiolog\'ia (CAB) CSIC-INTA, Camino Bajo del Castillo s/n, 28692, Villanueva de la Cañada, Madrid, Spain \label{aff:cab}
              \and
              European Space Agency (ESA), European Space Astronomy Centre (ESAC), Camino Bajo del Castillo s/n, 28692 Villanueva de la Cañada, Madrid, Spain \label{aff:esa}
              \and
              Nicolaus Copernicus Astronomical Center, Polish Academy of Sciences, ul. Bartycka 18, 00-716, Warsaw, Poland \label{aff:camk}
             }

    \date{Received XXXX; accepted YYYY}

 
\abstract
    {As stars orbit the Galaxy, interactions with structures such as the bar, spiral arms, and giant molecular clouds reshape their trajectories. While the effects on stellar orbits are well known, the impact on planetary systems remains largely unexplored.}
    {We investigate whether Galactic stellar motion leaves observable imprints on planet-hosting stars currently in the solar vicinity. We quantify radial mixing in exoplanet hosts and explore links between birth environment, orbital evolution, planetary architecture, and Galactic habitability.}
    {We constructed a homogeneous catalogue by cross-matching the Encyclopaedia of Exoplanetary Systems with \gaia\ DR3 astrometry and infrared photometry from 2MASS and AllWISE. Stellar orbits were integrated using \textsc{Galpy}. Stellar birth radii were inferred by combining Galactic chemical enrichment models with the generalised additive model introduced in \citetalias{Dantas2025a}.}
    {Most planet-hosting stars in our sample likely formed at smaller Galactocentric radii than their current guiding radii. Giant-planet hosts preferentially trace inner-Galaxy birth sites, whereas brown-dwarf hosts span a broader, less localised range of radial displacements. Rocky-only systems show smaller radial excursions and less centrally concentrated birth radii, while rocky+giant systems are intermediate, retaining a stronger link to inner-disc birth environments than rocky-only systems. Some outer-Galaxy-born hosts show signs of dynamical perturbation, especially in their vertical orbital structure. This is partly tied to the intermediate thin–thick disc population and may reflect age-dependent vertical heating or external perturbations. We also find that outward-migrators host more compact outer detected companions than inward-migrators, with non-migrators in between. This trend remains tentative because of heterogeneous detection biases.}
    {Giant-planet hosts retain a strong connection to metal-rich inner-Galaxy birth environments, whereas brown-dwarf hosts span a broader range of radial displacements, and rocky-only systems are less centrally concentrated. The older ages of rocky and rocky+giant hosts, especially among outward migrators, make them useful reference populations for future habitability and technosignature searches. Dynamically heated outer-Galaxy-born hosts show that planet-hosting systems can survive significant Galactic perturbations, although whether their architectures retain causal imprints of this evolution remains uncertain. No clear connection is found between radial displacement and the number of detected planets.}


   \maketitle
   \nolinenumbers  

\section{Introduction}
\label{sec:intro}

Galaxies are dynamic ecosystems shaped by the interplay of stars, gas, dust, and dark matter over time; the Milky Way (MW) is no exception \citep[e.g.][]{Minchev2013, Bland-HawthornGerhard2016, Matteucci2021}. Stars form from collapsing molecular clouds, with their masses following the initial mass function \citep[IMF; e.g.][]{Salpeter1955, Kroupa2001, Chabrier2003}. Recent work on the physical origin of the IMF reveals how the low-mass end is regulated by turbulence and feedback processes, linking star formation physics to the abundance of the smallest mass objects, including planetary-mass bodies \citep[e.g.][]{Gennaro2020, Hennebelle2024}. These planets, occupying the lowest mass end of the IMF, are not mere by-products of star formation, but active participants in Galactic evolution. Their formation, stability (or lack thereof), and dynamics are intimately linked to the life cycles of their host stars and to the broader Galactic environment \citep[e.g.][]{Haywood_2009, Adibekyan2014, Boettner2024}.

In parallel, evidence from both observations and simulations indicates that the MW's large-scale structures (such as the bar, spiral arms, and giant molecular clouds) gradually exert gravitational forces that perturb stellar orbits and, by extension, the planetary systems they potentially host \citep[e.g.][]{Castro1997, Brunetti2011, Martinez-Medina2016, MartinezBautista2021, Spitoni2025}. These perturbations can take different forms: (i) churning (also known as radial migration), where stars change angular momentum, often leading to more circularised orbits; (ii) blurring, a form of kinematic heating that does not produce net changes in angular momentum \citep[see, for instance,][and references therein]{SellwoodBinney2002, Lepine2003, Roskar2008, Kordopatis2015a, Halle2015, Chen2019, Feltzing2020, Khoperskov2020a, Carr2022, Dantas_2023, Iles2024, Nepal2024}; or (iii) minimal disturbance (or complete lack thereof), as seen in young Galactic components, such as open clusters and Cepheids, which have not yet undergone significant gravitational interactions with Galactic structures \citep{daSilvaDorazi2023, ViscasillasVazquez2023}. While these processes are well characterised in Galactic dynamics, their planetary consequences remain poorly understood, particularly whether they influence planet formation efficiency, system architectures or even promote planetary disconnection.

From the planetary-system perspective, the ever-growing exoplanet census offers an unprecedented window into how a star’s Galactic environment can influence planetary evolution. Most planet-hosting stars form in dense clusters \citep[e.g.][]{Hao2013}, often near massive OB-type stars whose intense ultraviolet (UV) can photo-evaporate nearby protoplanetary discs, altering their mass and geometry, thereby reshaping planet-formation pathways and long-term system survival \citep[e.g.][]{Adams2004, Winter2022, Berne2024, Coleman2025}. Young FGKM-type stars can likewise emit substantial UV flux during early evolutionary stages, further contributing to disc dispersal within their own systems \citep[e.g.][]{Shkolnik2014, Richey-Yowell2023, Li2025}.

While such `local cluster-scale effects' are relatively well established, a star’s broader Galactic trajectory may be as important. Beyond the well-known correlation between planet occurrence and stellar metallicity \citep[e.g.][]{FischerValenti2005, Johnson2010, Mortier2012, Teske2024}, the role of churning and blurring in shaping planetary architectures remains unclear. As stars migrate inward or outward, they may traverse regions of enhanced interstellar density, stronger radiation fields, or higher encounter rates; or escape them altogether \citep[see e.g.][for environmental processing in clustered birth environments and external irradiation effects]{Parker2020, Winter2018}. Such environmental changes can influence disc evolution and the long-term stability of planetary systems \citep[see e.g.][for disc dispersal, dynamical encounters, and planetary-system survival in dense stellar environments]{Adams2004, Fujii2019}, potentially increasing the production of free-floating planets \citep[e.g.][]{vanElteren2019}. In this sense, migration may act as a long-term modulator of the initial environmental imprint set during a star’s birth \citep[see e.g.][for Gyr-scale perturbations from Galactic tides, stellar flybys, and their effects on planetary architectures]{Kaib2013, Brown2022}.

These environmental and dynamical processes converge in the emerging framework of Galactic habitable zones (GHZs; covering both the MW and other galaxies), first introduced by \citet{Gonzalez2001}. Though still an incipient and rapidly evolving subfield of astrophysics \citep[e.g.][]{Stojkovic2019, Baba_2024}, the GHZ concept provides a natural framework for incorporating the effects of Galactic dynamics---including radial migration---into discussions of planetary habitability. Recent models that combine detailed MW chemical evolution histories with a parametric description of radial mixing suggest that stellar redistribution can enlarge the pool of stars capable of hosting habitable planets in the outer disc; by contrast, in the inner Galaxy, the presence of gas giant companions appears to boost the probability of forming rocky, potentially habitable worlds \citep[see][]{Spitoni2025}.

The Solar System itself offers a compelling test case for how radial migration may influence planetary habitability. Multiple lines of evidence indicate that the Sun was born in the inner Galaxy and later migrated outward to its present location \citep{Minchev2013, Minchev2018, Frankel2018, Baba_2023, Lu2024, Dantas2025a} at $\sim$ 8.2 kpc \citep{McMillan2017}. Along this journey, Earth’s climate may have been perturbed on several occasions, particularly during passages through the Galactic spiral arms \citep{Tsujimoto_2020, BJ2022}. Such episodes highlight that habitability can be fragile and contingent: planets may remain clement for long intervals yet still be vulnerable to stochastic hazards. Indeed, \citet{Tyrell2020} has shown through simulations that chance played a decisive role in maintaining Earth’s habitability across Gyr timescales. In that context, it is sobering that, on timescales orders of magnitude shorter, anthropogenic greenhouse gas forcing is now demonstrably shifting Earth’s climate baseline, adding a distinctly human source of instability to a planet whose long-lived habitability is not guaranteed \citep[e.g.][]{Haustein2017, Ribes2021}; similarly, large-scale human conflict can erode the social and environmental stability on which long-term habitability ultimately depends \citep[e.g.][]{Spiegel2024, Marcantonio2025}.

Taken together, these insights point to a continuous interplay between birth-environment effects and long-term Galactic evolution. The former sets the initial conditions for protoplanetary disc properties through processes such as UV photo-evaporation, external irradiation, and dynamical encounters in dense stellar environments \citep[e.g.][]{Adams2004, Winter2018, Winter2022}. The latter redistributes planetary systems over Gyr timescales through churning and blurring \citep[e.g.][]{SellwoodBinney2002, Chen2019, Chen2020}. As a consequence, planet-hosting stars may be carried across Galactic regions that differ in FUV radiation field, stellar density, ISM properties, and dynamical hazard \citep[e.g.][]{Winter2019, Kaib2013, Spitoni2025}. Both processes therefore modulate the extent and structure of the GHZ, influencing where life-friendly planets are most likely to arise and persist in the Galaxy \citep[e.g.][]{Gonzalez2001, Spitoni2025}.

Building on this framework, we adopt the technique developed by \citet[][hereafter \citetalias{Dantas2025a}]{Dantas2025a}, which employs a generalised additive model (GAM) to extend the chemical enrichment models originally developed by \citet{Magrini2009}. This approach allowed us to infer stellar birth radii ($R_b$) for a curated \gaia-ESO sample of thin disc stars \citep{Randich2022, Gilmore2022}. In the second paper of this series \citep[][hereafter \citetalias{Dantas2025b}]{Dantas2025b}, we tested the hypothesis that radial migration could enhance lithium (Li) depletion in the same sample as \citetalias{Dantas2025a} using a survival analysis strategy \citep[a brief extension to the Solar case is presented in][]{Dantas2025c}. Such hypothesis had been previously raised by \citet{Guiglion2019} and further explored by \citet{Dantas2022}, but ultimately discarded; although migration classes exhibit apparent differences in Li abundance and signs of depletion, \citetalias{Dantas2025b} showed that these trends are explained by stellar intrinsic properties rather than by migration itself.

In this third paper, we use the \citetalias{Dantas2025a} $R_b$ inference to test whether stellar migration histories are reflected in planetary architectures and survival. By linking stellar birth radii to present-day systems, we provide a direct empirical probe of how radial migration shapes exoplanet demographics across the Galaxy, and how similar processes may operate in other MW-like galaxies. More broadly, this work underscores how Galactic evolution influences planetary survival and potential habitability.

This paper is structured as follows. In Sect. \ref{sec:data_method} the data and methodology are presented; in Sect. \ref{sec:analysis_discussion} we present our results, analyses, and discussion; in Sect. \ref{sec:conc} we present our conclusions. The data availability is presented at the end of this manuscript.

\section{Data and methodology}
\label{sec:data_method}

\subsection{Data curation}
\label{subsec:catalogue}

\subsubsection{Planetary systems catalogue}
\label{subsubsec:data_planets}

The basis of our study is the \textit{Encyclopaedia of Exoplanetary Systems} catalogue\footnote{Available at \url{https://exoplanet.eu/home/}.} (hereafter `\texttt{exoplanets} catalogue'), a comprehensive and continuously updated database curated by the astrophysical community. We selected entries flagged as `confirmed' and, following the catalogue definition, retained objects with reported masses below 60 Jupiter masses \citep[\mjup;][]{Schneider2011}.

This pragmatic upper bound includes the brown-dwarf (BD) regime while remaining safely below the canonical hydrogen-burning limit of $\sim$75--80 \mjup, whose exact value depends on composition and evolutionary modelling \citep[see e.g.][]{ChabrierBaraffe2000, Burrows2001}. It therefore excludes bona fide stellar companions. The adopted limit also reflects the empirical continuity of the observed mass--radius--density locus across the high-mass end of the giant-planet population \citep{Hatzes2015}, while accommodating cases where quoted 1$\sigma$ uncertainties would otherwise move an object across a sharper threshold. This convention avoids imposing a sharp planet--BD boundary, particularly because the deuterium-burning threshold near 13 \mjup\ is not a unique separator in practice \citep{Spiegel2011}. Where relevant, however, we stratify the analysis by mass regime, distinguishing between planetary and BD systems.

Available planetary parameters in the \texttt{exoplanets} catalogue include masses, radii, orbital periods, semi-major axes (\amax), eccentricities, inclinations, and detection methods. Stellar properties---such as positions, apparent magnitudes, metallicities, masses, radii, and effective temperatures (\teff)---are primarily retrieved from SIMBAD \citep{Wenger2000} and the published literature. Despite unavoidable heterogeneity across surveys, the catalogue provides sufficient metadata to assess the provenance and reliability of individual measurements.

The \texttt{exoplanets} catalogue, restricted to confirmed detections\footnote{Updated as of 7 November 2024.}, was cross-matched with \gaia\ data release 3 \citep[\gaia\ DR3;][]{Gaia2023_DR3} using DR3 source identifiers retrieved from SIMBAD. The resulting dataset combines planetary parameters (e.g. masses, radii, and orbital periods) with host-star astrometric information, including parallaxes and proper motions, as well as additional stellar parameters provided by \gaia.

\subsubsection{Host-star data and quality cuts}
\label{subsubsec:data_stellar}

To minimise contamination from potentially perturbed astrometric solutions, we retained only sources satisfying the \gaia\ DR3 quality criteria \texttt{ruwe}$<1.4$, \texttt{ipd\_frac\_multi\_peak}$\leq2$, and \texttt{ipd\_gof\_harmonic\_amplitude}$<0.1$, which are commonly used to identify problematic astrometric fits and possible unresolved multiplicity \citep[e.g.][]{Fabricius2021, Kervella2022}. We also removed sources with unphysical parallax solutions. Infrared photometry was retrieved from 2MASS \citep[$JHK_s$;][]{2MASS} and AllWISE \citep[$W1W2$;][]{AllWise} using \textsc{TopCat} \citep{TOPCAT}, adopting a conservative 2 arcsec matching radius.

Finally, we retrieved \feh\ values from the \texttt{SWEET-Cat} catalogue \citep{Santos2013}, as it is critical for \rbirth\footnote{We adopt median values hereafter, due to a resampling procedure described in \citetalias{Dantas2025a}.} estimation for our stellar sample, which is detailed in Sect. \ref{subsec:orbits_and_rb}. For one star in the sample (\texttt{ID}=4794830231453653888), multiple \feh\ estimates were available; in this case, we conservatively retained the value with the largest uncertainty, which encompasses the alternative measurement.

\subsection{Orbit integration and birth radii estimation}
\label{subsec:orbits_and_rb}

We model the Galactic orbits of all planet-hosting stars in our sample to assess how stellar motion within the MW may have influenced the architecture and long-term evolution of their planetary systems. Orbit integration traces stellar trajectories within a prescribed Galactic potential over Gyr timescales, allowing us to characterise each star’s dynamical behaviour and to infer key quantities relevant to its radial evolution within the disc (e.g. guiding radii, angular momentum, and action components, eccentricities).

Using \gaia\ DR3 radial velocities and astrometric quantities (parallaxes, proper motions, and their respective uncertainties), we first applied the parallax zero-point correction for sources with either \texttt{pseudocolour=NAN} or $1.24 \leq \rm{\texttt{pseudocolour}} \leq 1.72$, and \texttt{astrometric\_params\_solved} $>3$ \citep[as prescribed by][]{Lindegren_2021}. Corrected Bayesian distances were then derived following the prescription of \citet{Bailer-Jones_2015}.

Following \citet[][Sect. 2.2]{Dantas_2023}, we performed a bootstrapping analysis to propagate observational uncertainties into distributions of the dynamical parameters prior to orbit integration. From these distributions, we derived percentile-based confidence intervals (including the median, 50th percentile) for all relevant quantities. For practical purposes, and unless otherwise stated, we adopt the median values of these distributions as representative inputs for the orbit integration (e.g. median guiding radius, \rgui). 

Using the resulting bootstrapped parameter sets, we integrated the Galactic orbits of all planet-hosting stars with \textsc{Galpy} \citep{Bovy2015}. Orbits were computed over a 10 Gyr lookback time within the \citet{McMillan2017} MW potential, using the Dormand--Prince integration scheme implemented in \textsc{c} (i.e. \texttt{method=\textquotesingle dop853\_c\textquotesingle}; \citealt{DormandPrince1980}).

In addition to the dynamical analysis, we also require $R_{b}$ estimates for our stellar sample. For this, we followed the prescription from \citetalias{Dantas2025a}, where \rbirth\ is derived by making use of a GAM \citep[][]{HastieTibshirani1990} to extend \citet{Magrini2009}'s Galactic chemical enrichment models. While \citet{Magrini2009}'s gradients incorporate [O/Fe], [Si/Fe], [Mg/Fe], [Ca/Fe], and \feh, the method presented in \citetalias{Dantas2025a} relies on a minimalist approach which uses only \feh\ and stellar age (\age), parameters that are available for all the stars in our planet-hosting stars catalogue; the other abundance ratios are highly correlated with \feh, which are redundant for the GAM. It is worth mentioning that other works rely on a similar approach, such as the one by \citet{Ratcliffe2023}, who also only make use of \feh\ and stellar age (\age). For details on \rbirth\ estimation via the GAM, as well as its strengths and limitations, we refer the interested reader to \citetalias{Dantas2025a}.

\subsection{Galactic component classification of the stellar sample}
\label{subsec:galactic_components}

We classified the host stars according to their likely Galactic component. This step is necessary because the chemical evolution models adopted to infer \rbirth\ are calibrated for the Galactic thin disc, where radial metallicity gradients are well defined \citep{Magrini2009}. Conversely, thick disc and halo stars may follow different enrichment histories, and halo stars in particular may include objects of accreted or extragalactic origin (e.g. Gaia--Enceladus; \citealt{Helmi2008, Bonaca2017}). We therefore retained only stars classified as thin disc or intermediate objects, excluding likely thick disc and halo contaminants. We refer the reader to \citetalias{Dantas2025a} for a longer discussion on the chemical enrichments of each Galactic component.

\begin{figure*}
   \centering
   \includegraphics[width=\linewidth]{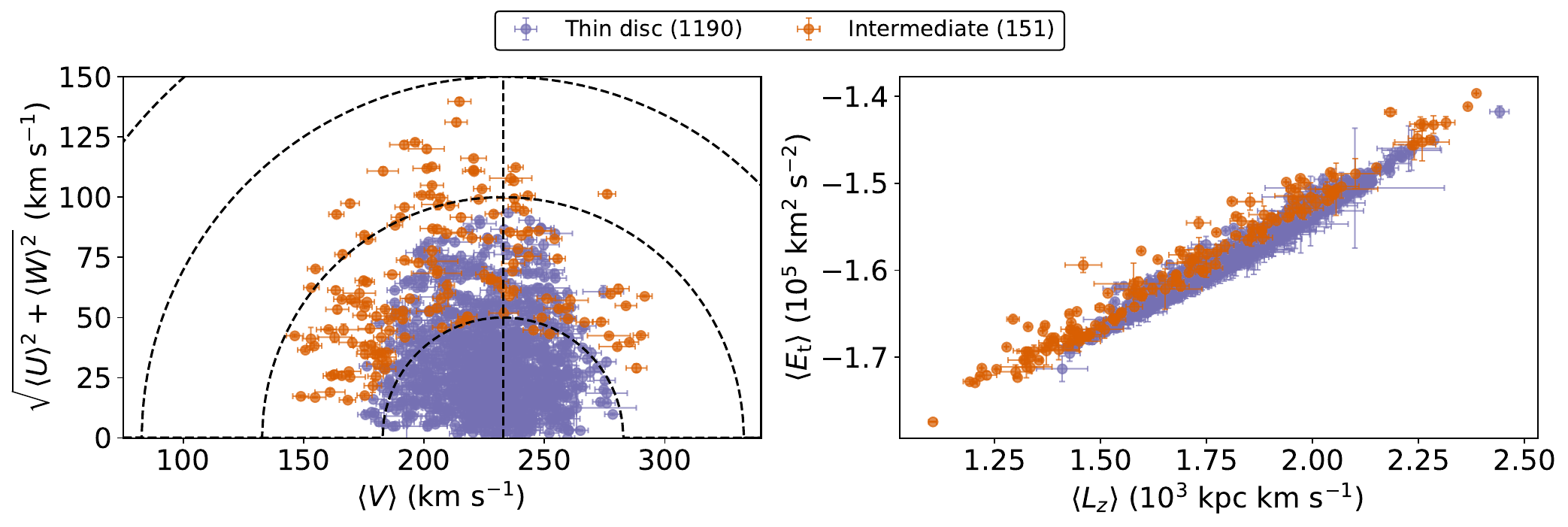}
   \caption{Toomre and Lindblad diagrams for our sample. Left panel: Toomre diagram depicting the classification for the stars in our sample based on their dynamics, according to the prescription of \citet{Bensby_2003}. Thin disc members are depicted in purple, and intermediate stars in orange markers. Right panel: Lindblad diagram showcasing our sample as classified in the Toomre diagram.}
   \label{fig:Toomre_Lindblad}
\end{figure*}

To classify the planet-hosting stars of our sample as members of the different Galactic components, we make use of their kinematic signatures. We adopt the prescription of \citet{Bensby_2003}, using Galactic space velocities ($U$, $V$, $W$) derived from the orbit integration. This method accounts for the relative local densities of the Galactic components, reflecting the higher probability of sampling thin disc stars in the solar neighbourhood. For each star, we compute membership probabilities and derive the thick-to-thin disc (TD/D) and thick-to-halo (TD/H) likelihood ratios. Stars with TD/D $< 0.1$ are classified as thin disc members, those with TD/D $> 10$ as thick-disc members, and intermediate values are assigned to a transition population. Halo candidates are identified from their low TD/H likelihood ratios. We therefore retain only stars classified as thin disc or intermediate thin--thick disc objects for the subsequent \rbirth\ analysis, thereby maximising the usable sample. Figure \ref{fig:Toomre_Lindblad} presents the resulting Toomre and Lindblad diagrams.

\subsection{The final stellar sample}
\label{subsec:final_sample}

The final sample comprises 1341 planet-hosting stars within $\sim$ 2.3 kpc, of which 1190 belong to the Galactic thin disc and 151 are classified as the intermediate population. Our analysis is focused hereafter on the thin disc members, for which the adopted metallicity gradient models were developed \citep{Magrini2009}; and, given the known overlap between the thin and thick discs \citep[see the discussion developed in][to mention a few]{Bensby2014, Nidever2014, Kordopatis2015b}, we also include the intermediate stars, as they could potentially be thin disc members. The distributions of \feh, \age, stellar mass ($M_{\star}$), and \teff\ are shown in Fig. \ref{fig:combined_distributions}, while their complementary spatial distribution across the MW is displayed in Fig. \ref{fig:sky_coverage_MW_map} in Appendix \ref{appendix:subsec_stellarprop}. The median distances are $\sim$ 380 pc; we display the heliocentric distances in Fig. \ref{fig:heliocentric_distances} also in Appendix \ref{appendix:subsec_stellarprop}. 

\begin{figure}[!ht]
   \centering
   \includegraphics[width=\linewidth]{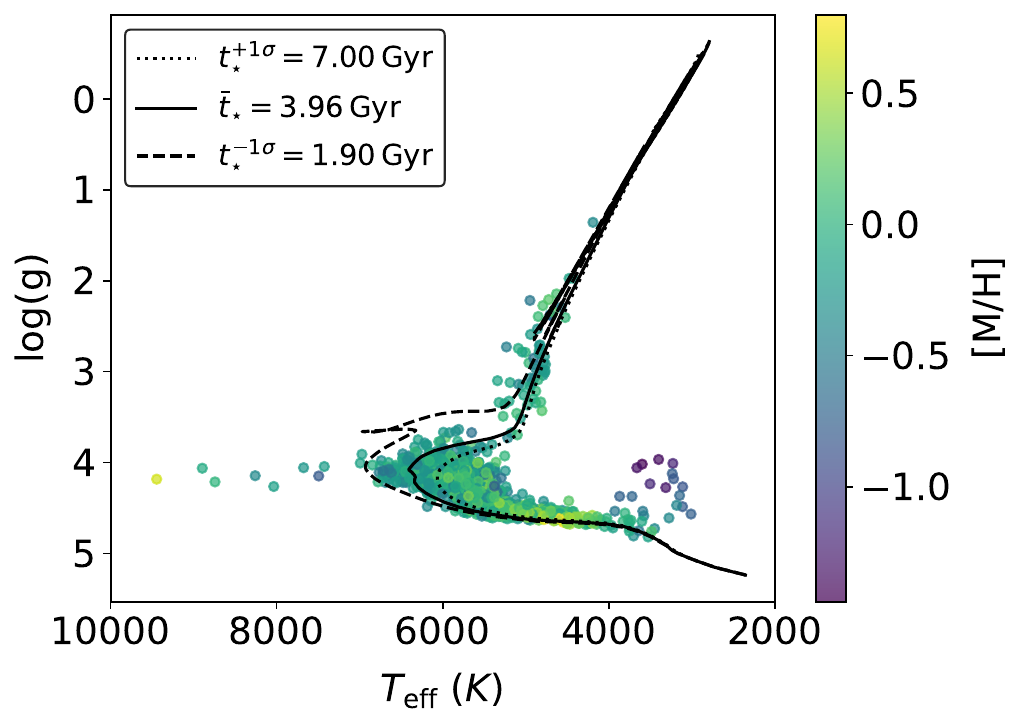}
   \caption{Kiel diagram depicting PARSEC isochrones \citep{Bressan2012} for the median [M/H] for our sample stratified by age (median value and a $\pm 1 \sigma$ variation shown by the continuous, and dashed and dotted tracks, respectively). The colour-map shows the variation of [M/H] for the stars in our sample.}
   \label{fig:kiel}
\end{figure}

The Kiel diagram of our sample (Fig. \ref{fig:kiel}) shows that the bulk of stars are of FGK-type with some potential A-type interlopers, with a median \teff\ of $\sim$ 5600 K, which is very close to the Sun’s effective temperature ($T_{\rm eff,\odot} = 5772.0 \pm 0.8$~K; \citealt{Prvsa2016}, which is also consistent with \citealt{Asplund2021}). The other median parameters of our sample are likewise close to solar values: $\langle \rm{\feh} \rangle = 0$ \citep[by definition;][but see also \citealt{Korn2003} for a discussion on the non-local thermodynamic equilibrium corrections for \feh\ in the Sun]{Asplund2021} and $t_{\odot} = 4.775 \pm 0.039$ Gyr \citep[][]{BonannoFrohlich2015}; see Fig. \ref{fig:combined_distributions}.

\subsection{Stellar motion classification}
\label{subsec:stellar_migration}

To quantify the radial displacement of stars in our sample, we follow the procedure described in \citetalias{Dantas2025a}, which combines the uncertainties on \rbirth\ and guiding radii (\rgui) to assess whether a statistically significant radial shift has occurred. Stars for which \rbirth\ and \rgui\ are consistent within 2$\sigma$ are classified as non-migrators and labelled as \texttt{equal}. Stars with differences exceeding 2$\sigma$ are classified as having migrated either \texttt{inward} (\rgui $<$ \rbirth) or \texttt{outward} (\rgui $>$ \rbirth).

We do not attempt to distinguish between stars that remain undisturbed and those that have experienced blurring, as both scenarios conserve angular momentum. Systems classified as \texttt{equal} (which we also refer to as `non-migrators') may therefore include both genuinely non-migrators and older stars whose orbital radii have remained unchanged despite dynamical heating \citep[i.e. blurred; see e.g.][]{Feltzing2020}. Further details and a comprehensive discussion of this classification scheme are provided in \citetalias[][Sect.~3.3 therein]{Dantas2025a}.

\subsection{Planet type classification and sample properties}
\label{subsec:planet_type}

We classify the companions in our sample into three broad categories: rocky planets, giant planets, and BDs. The primary division between planetary and substellar companions is set at 13~\mjup, which we adopt as the conventional boundary between planets and BDs \citep{Spiegel2011}.

Below this threshold, planetary companions are further separated into low-mass and giant planets. Objects with masses below 10 Earth masses ($M_{\oplus}$)---corresponding to the super-Earth regime---are classified as `rocky' planets, while more massive companions are grouped as `giant' planets.

For planets lacking direct mass measurements, we estimate masses from the observed radii in order to maintain a homogeneous classification across the sample. The consensus in the literature \citep[e.g.][]{WeissMarcy14} is that objects with $R < 1.6~R_{\oplus}$ are considered rocky and we can therefore assume a bulk density of 5.5~g~cm$^{-3}$, comparable to that of the Earth; while for larger-radius objects we should adopt a representative density of 1~g~cm$^{-3}$, typical of gaseous planets in the Solar System. These assumptions yield approximate mass estimates that allow us to place such objects within the same mass-based categories as planets with measured masses. As a sanity check, we compared our results with a sample where a limit of $R < 2.2~R_{\oplus}$ for rocky planets was set, without any significant change in the conclusions reached. 

Having defined companion categories, we then classify the host stars according to the types of companions they harbour. All stars in our sample host at least one confirmed companion and are assigned to one of the following system-level categories: systems hosting only giant planets (966 stars), both giant planets and BDs (16), only BDs (13), both rocky and giant planets (93), or only rocky planets (253).

This classification scheme is designed to provide a uniform, population-level framework for comparing planetary architectures across the sample. It is not intended for detailed compositional or dynamical characterisation of individual systems, but rather to enable robust statistical comparisons in the context of stellar motion and Galactic environment.

\section{Results, analysis, and discussion}
\label{sec:analysis_discussion}

\subsection{Radial migration diagnostics}
\label{subsec:diagnostics}

In this section, we assess radial migration using two complementary approaches: (i) a direct comparison between \rbirth\ and \rgui; and (ii) an analysis of additional dynamical parameters commonly associated with radial migration or lack thereof.

\subsubsection{Birth versus guiding radius offsets as a migration diagnostic}
\label{subsubsec:rb_versus_rg}

\begin{figure*}
    \centering
    \includegraphics[width=\linewidth]{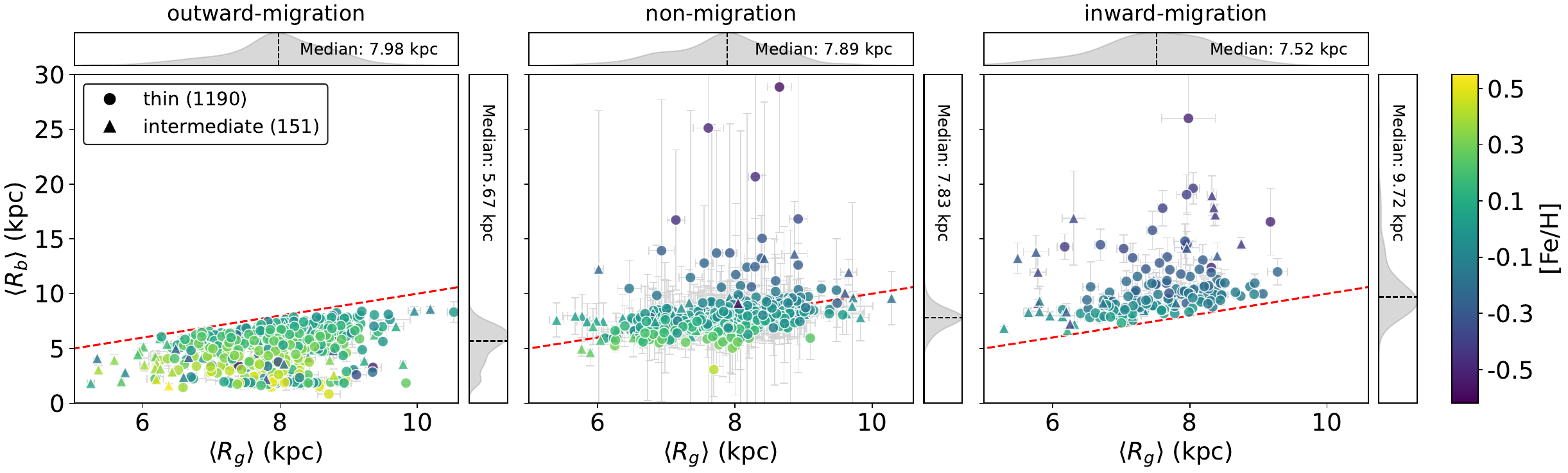}
    \caption{Comparison between estimated Galactocentric birth radii (\rbirth) and present guiding radii (\rgui) for stars in the thin disc and intermediate populations (see classification in Sect. \ref{subsec:galactic_components}). The three panels correspond to outward-migrators (left), non-migrators (middle), and inward-migrators (right). Points are colour-coded by metallicity (\feh) using a continuous colour map, while marker shape indicates to which Galactic disc component they most likely belong: circles denote thin disc stars, while triangles correspond to the intermediate population. The red dashed line marks the one-to-one relation between \rbirth\ and \rgui. The mean metallicities are $\langle \rm{\feh} \rangle = 0.17$, $0.01$, and $-0.17$ for the outward, equal, and inward categories, respectively. Adjacent plots show the distributions for \rbirth\ and \rgui\ with medians annotated in their respective sub‑panels.}
    \label{fig:birth_guiding_radii_plot}
\end{figure*}

\begin{figure*}
   \centering
   \includegraphics[width=\linewidth]{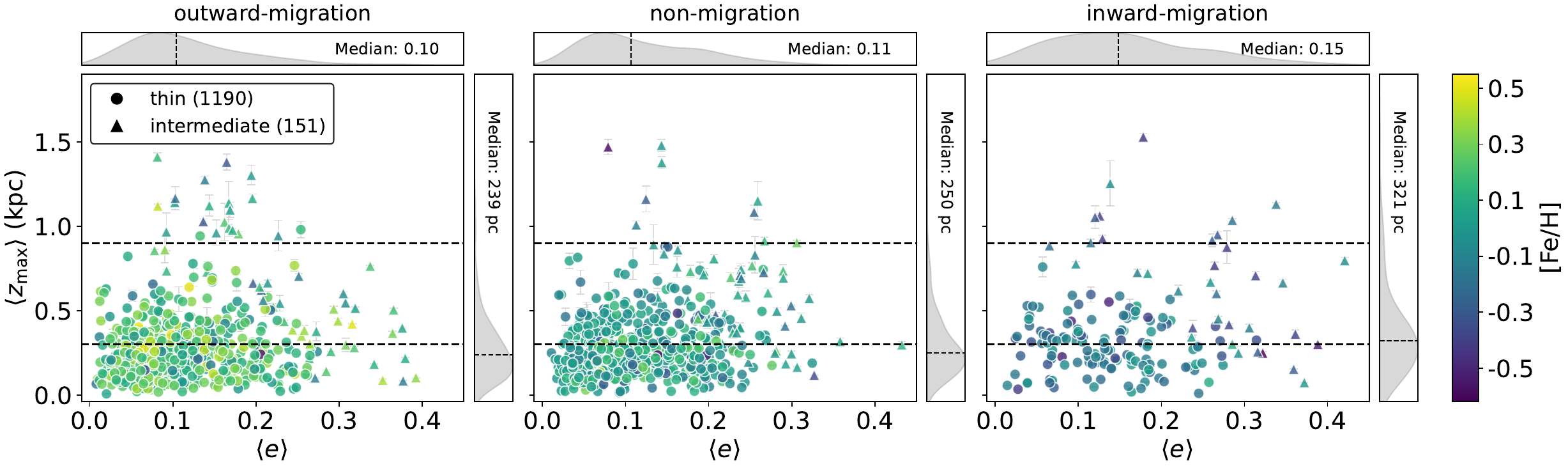}
   \caption{Median maximum Galactic scale heights (\zmax) versus median orbital eccentricity (\eccentricity), split by kinematic class (panels), Galactic component (marker shape), and \feh\ (colour map). From left to right the panels show outward‑, non‑, and inward‑migrators. Dashed horizontal lines indicate the canonical thin disc ($Z_{\rm thin}=300\,{\rm pc}$) and thick disc ($Z_{\rm thick}=900\,{\rm pc}$) scale-heights. Each main panel is accompanied by inset sub‑panels that display the full distributions of \eccentricity\ and \zmax\ for the corresponding subsample, with the median values marked.}
   \label{fig:zmax_vs_eccentricity}
\end{figure*}

Figure~\ref{fig:birth_guiding_radii_plot} shows \rbirth\ versus \rgui, colour-coded by \feh\ and stratified into stars classified as outward-migrators (left panel), non-migrators (middle panel), and inward-migrators (right panel). The metallicity gradients follow the expected trend, consistent with inside-out disc formation and growth scenarios \citep[e.g.][]{Matteucci1989, Chiappini1997, Kobayashi2006, Bergemann2014, Schonrich2017, Hu2023, Magrini2023}: stars formed in the inner Galaxy exhibit higher \feh, whereas those born at larger Galactocentric radii display lower metallicities. 

As expected systems formed closer to the inner Galaxy tend to move outward, whereas those formed at larger radii tend to migrate inward; non-migrators occupy intermediate formation radii. This behaviour is in agreement with previous results \citep[e.g.][as well as \citetalias{Dantas2025a}]{Martinez-Medina2016, Dantas_2023}. In contrast, the \rgui\ values, representing the current average orbital radii, are considerably more mixed within each stratified subclass of motion and planet type.

Figure \ref{fig:migration_distr} in the Appendix complements this view by showing the distribution of $\Delta R \equiv$ \rgui - \rbirth. The distribution indicates that the majority of stars in our sample are consistent with having formed at smaller Galactocentric radii than their present \rgui, with a smaller fraction formed at larger radii and now observed in the solar neighbourhood. Despite the different nature of the present sample (restricted to stars hosting planetary or substellar companions) this qualitative behaviour closely resembles that reported in \citetalias{Dantas2025a}, where \gaia-ESO field stars were analysed using the same \rbirth\ framework. In all cases, outward-migrators from the inner disc dominate the local migrators' population, with those moving inward constituting a smaller contribution.

Stars with very large inferred birth radii (i.e. $\gtrsim$ 20–30 kpc) may either correspond to interlopers from the thick disc (an interpretation supported by their inclusion among the intermediate populations selected in the Toomre diagram and indicated by triangle-shaped markers) or reflect limitations of the model when confronted with specific age–metallicity combinations. This effect is most apparent among stars classified as inward-migrators, for which a subset of large \rbirth\ objects is plausibly associated with the thick disc. Other stars with large \rbirth\ estimates typically carry substantial uncertainties; as they represent only a small fraction of the sample, they do not affect the global trends discussed here.

We do not intend to revisit all the strengths and caveats of the GAM applied to the metallicity gradient models of \citet{Magrini2009}, which are employed in the current manuscript; instead, we refer the reader to \citetalias{Dantas2025a} for a detailed discussion, where, for instance, Appendix A describes the GAM in depth, Sect. 3.2.1 discusses some of its features and limitations, and Fig. C8 illustrates the model’s reduced reliability for stars with certain metallicities formed at very small radii, close to the Galactic bulge. Additional relevant discussions on the model and its applications are provided throughout that paper.

\subsubsection{Additional dynamical diagnostics of radial migration}
\label{subsubsec:additional_diagnostics}

Figure~\ref{fig:zmax_vs_eccentricity} provides a complementary view of radial migration through two additional dynamical parameters: the maximum vertical height (\zmax) and orbital eccentricity (\eccentricity). Both have been widely discussed in the context of migration, especially in MW-like simulations. Variations in \eccentricity\ are commonly interpreted as tracers of orbital heating \citep[e.g.][]{Khoperskov2020a}, while \zmax\ is predicted to increase preferentially for outward-migrators, which move into regions of lower vertical restoring force and can therefore reach larger vertical amplitudes \citep{Roskar2012}. For reference, the horizontal dashed lines in Fig.~\ref{fig:zmax_vs_eccentricity} mark the adopted thin and thick disc scale-heights (300 and 900 pc, respectively; \citealt{McMillan2017}).

All stars in our sample exhibit near-circular orbits, with median \eccentricity\ values between 0.11 and 0.15 and maximum values of order $\sim 0.4$ across migration classes (Fig.~\ref{fig:zmax_vs_eccentricity}). This is consistent with observational studies of the solar neighbourhood, where thin-disc stars typically occupy low-\eccentricity\ orbits, peaking around \eccentricity~$\sim$0.1--0.2 and showing only rare higher-\eccentricity\ outliers \citep[e.g.][]{Kordopatis2011, Kordopatis2015a, Chen2019}. Such near-circular orbits are also expected for cold thin-disc tracers with small asymmetric drift \citep[e.g.][]{Huang2015}. Similarly, \citetalias{Dantas2025a} found that \eccentricity\ did not meaningfully distinguish between migrators and non-migrators, as that sample displayed uniformly low values. This contrasts with some recent theoretical expectations predicting systematically higher eccentricities for blurred or non-migrating stars \citep[e.g.][]{Lehmann2024}. Whether such uniformly low eccentricities are linked to the presence or detectability of planets remains unclear.

In terms of vertical structure, most stars in our sample remain confined to low \zmax\ values, consistent with thin-disc-like orbits. Nevertheless, a non-negligible fraction reaches heights more typical of dynamically heated populations, with the largest vertical excursions frequently associated with stars classified as intermediate thin--thick-disc objects. These elevated \zmax\ values are not restricted to a single motion class: they are present among outward-, non-, and inward-migrators. However, the inward-migrating subsample shows the clearest age--\zmax\ trend and includes several high-\zmax\ intermediate-population stars. We refer the reader to Fig. \ref{fig:zmax_age_components} in the Appendix, where these trends are further explored.

This suggests that the vertical signal is not simply a direct consequence of radial migration, but likely reflects a combination of residual population mixing, age-dependent heating, and additional perturbative processes. One possible contributor is the interaction between the MW and the Sagittarius dwarf galaxy, whose pericentric passage has been associated with vertical perturbations of the outer Galactic disc \citep[e.g.][]{Das2024}; however, this interpretation remains tentative given the large uncertainties in stellar ages and the probabilistic nature of the population classification.

Similar diversity is seen in simulations of MW analogues. \citet{Roskar2012} report enhanced vertical excursions for outward-migrators, whereas \citet{Minchev2012} argue that the effect should be weak or negligible, and \citet{VeraCiro2016} find that outward-migrators largely preserve their original scale-heights while inward-migrators become dynamically thinner. Our results for this sample of planet-hosting stars do not map cleanly onto any single one of these scenarios. Outward-migrators include stars with large \zmax, but comparable or larger vertical excursions also occur among non- and inward-migrators, especially within the intermediate population subset. In addition, the adjacent \zmax\ distributions shown in Fig. \ref{fig:zmax_vs_eccentricity} indicate a systematic increase in median \zmax\ values from outward-migrators to non-migrators and inward-migrators. With the present sample, we therefore cannot attribute the observed spread in \zmax\ to a single mechanism. Instead, it likely reflects a combination of migration-driven vertical heating, preservation of birth scale-heights, contamination by dynamically hotter populations, and perturbations unrelated to churning alone.

Overall, the behaviour of both \eccentricity\ and \zmax\ remains broadly consistent with the picture discussed in \citetalias[][]{Dantas2025a} and \citet{Dantas_2023}: \eccentricity\ offers limited discriminatory power between motion classes, while elevated \zmax\ values occur across multiple categories. The orbit integrations, however, show that dynamically heated outer-Galaxy-born and intermediate population planet hosts are present in the sample. This is qualitatively consistent with the trend found in \citetalias[][]{Dantas2025a}, where the most metal-poor thin-disc stars---which are preferentially inferred to have formed in the outer disc---also displayed larger vertical excursions. These results reinforce the difficulty of disentangling blurring from churning using \eccentricity\ alone, and suggest that radial displacement does not necessarily map onto a unique vertical heating signature. Instead, high-\zmax\ planet hosts likely trace a combination of radial displacement, vertical heating, and Galactic component mixing.

\subsection{Effects of stellar migration on planetary system properties}
\label{subsec:planetary_sys_prop}

When placing planetary systems in a Galactic dynamical context, a single kinematic or orbital diagnostic is insufficient. A meaningful characterisation requires a multi-dimensional approach that captures the chrono-chemo-dynamic properties of the host stars alongside the characteristics of their planetary companions. To this end, Fig. \ref{fig:heatmap} (presented in Appendix \ref{appendix:subsec:exoplanet_features}) shows Spearman rank correlations, visualised through a hierarchical clustering heatmap \citep[or `clustermap';][]{MurtaghContreras2012, Murtagh2014}, for a selected subset of stellar and planetary parameters. Strongly correlated quantities (particularly among orbit-integration outputs) were excluded for clarity, as they exhibit very similar behaviour and cluster tightly in the hierarchical representation. This selection highlights the dominant and most informative relationships.

\subsubsection{Planet formation within the Galaxy and role of metallicity}
\label{subsubsec:planet_galaxy_metallicity}

\begin{figure}
   \centering
   \includegraphics[width=\linewidth]{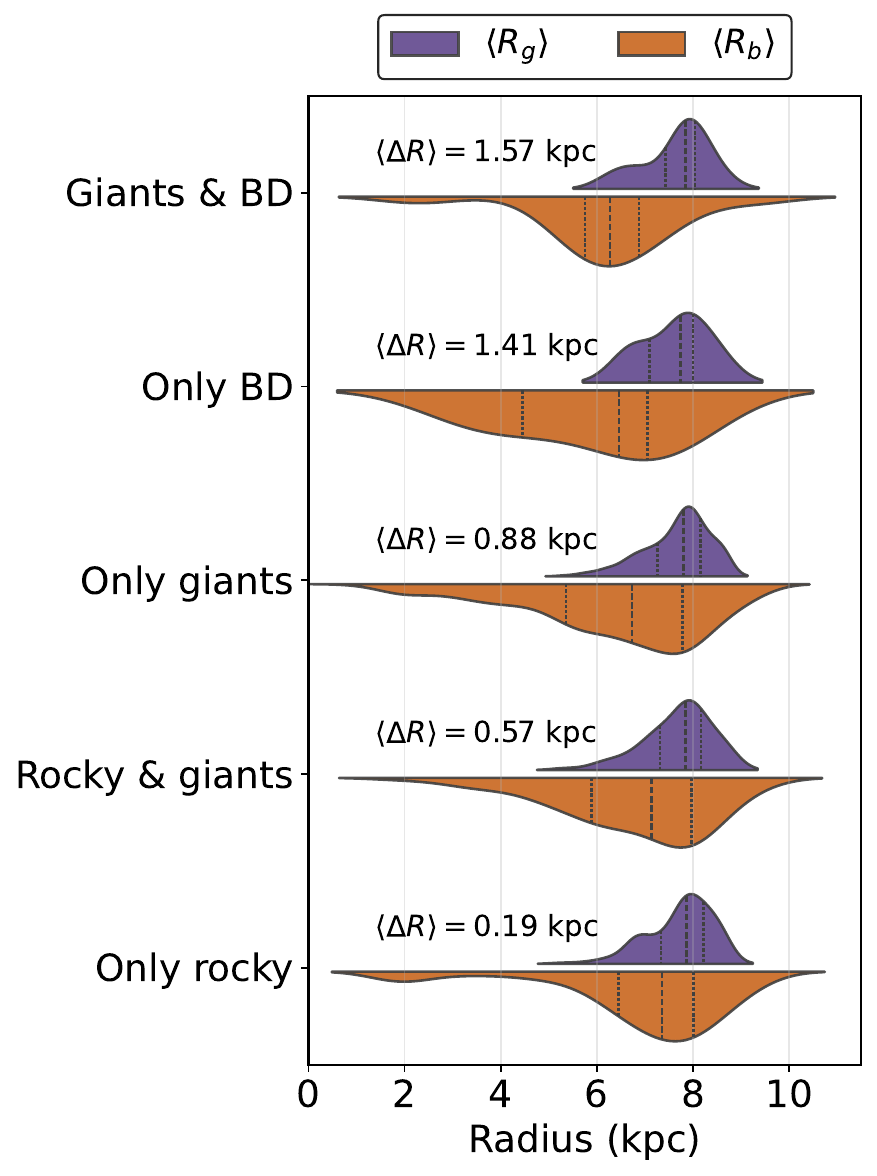}
   \caption{Distributions of guiding $\langle R_g \rangle$ (purple) and birth radii $\langle R_b \rangle$ (orange) for planetary systems grouped by planet-type category, with the median differences $\langle \Delta R \rangle$ annotated for each group, ordered accordingly. For visualization purposes only, the violin plots are truncated at the 90th percentile in radius to reduce the impact of extreme tails on the kernel density estimate. However, all median values are computed using the full, untrimmed sample.}
   \label{fig:violin_categories}
\end{figure}

The distributions show that the different planet-host classes do not occupy identical chemo-dynamical regimes. Systems hosting both giant planets and BDs have the smallest median inferred birth radius, with \rbirth$_{50\%}=6.3$ kpc, while BD-only systems show a similar median value, \rbirth$_{50\%}=6.5$ kpc, albeit with a broader dispersion (MAD $=1.2$ kpc; Table~\ref{tab:stats_rb_rg}). Giant-only systems occupy a slightly less centrally concentrated regime, with \rbirth$_{50\%}=7.0$ kpc, still below their median present-day guiding radius of \rgui$_{50\%}=7.9$ kpc. Mixed rocky+giant systems are intermediate, with \rbirth$_{50\%}=7.3$ kpc and \rgui$_{50\%}=7.9$ kpc. In contrast, rocky-only systems have the largest median birth radius, \rbirth$_{50\%}=7.8$ kpc, nearly matching their present-day guiding radius, \rgui$_{50\%}=8.0$ kpc. This sequence suggests that systems hosting giant planets and/or BDs retain a stronger connection to inner-disc birth environments than rocky-only systems, although the small BD-host samples require caution.

The BD-host populations add an important nuance to this picture. Although their median \rbirth\ values are comparatively small, they do not appear to follow the same sharply concentrated pattern as the giant-only hosts. The radial-displacement diagnostics in Table~\ref{tab:metallicity_age_deltar} are useful here: $W_{68}(\Delta R)$ traces the width of the dominant body of the distribution within each motion class bin, while $W_{90}(\Delta R)$ is more sensitive to extended tails. In particular, outward-migrating BD-only systems have $W_{68}(\Delta R)=3.57$ kpc, compared with $1.56$ kpc for systems hosting both giants and BDs. This suggests that BD-host stars may sample a broader, or less sharply localised, range of Galactic birth environments than giant-only hosts, for which the connection with metal-rich inner-disc formation is expected to be stronger \citep[in broad agreement with][]{Maldonado2017}.

These empirical findings can be interpreted in the context of the well-established correlation between stellar metallicity and giant-planet occurrence, whereby giant planets are more likely to form around metal-rich stars \citep[e.g.][]{Johnson2010, Mortier2012, Teske2024}. In the core-accretion framework, higher metallicity implies a higher dust-to-gas ratio and thus a larger solid surface density in the protoplanetary disc, accelerating core growth and increasing the likelihood of reaching the critical core mass for runaway gas accretion within the disc lifetime \citep[][]{Pollack1996, IdaLin2004}. The subsequent gas-accretion phase is regulated by envelope cooling and opacity, with lower grain opacities generally enabling faster contraction and more rapid gas accretion \citep[e.g.][]{HoriIkoma2010, Alessi2018}.

Within the framework of Galactic chemical evolution, metal-rich stars are expected to originate preferentially in the inner thin disc, where the interstellar medium experienced faster and more efficient enrichment \citep[e.g.][]{Chiappini1997, Kobayashi2006}. Indeed, several studies have identified metal-rich and super-metal-rich stars currently inhabiting the solar vicinity that exhibit chemo-dynamical signatures of an inner Galaxy origin \citep[e.g.][]{Castro1997, Pompeia2002, Trevisan2011, Kordopatis2015a, Chen2019, Zhang2021, Dantas_2023}. The Sun itself---a relatively metal-rich star---and the Solar System are likewise thought to have formed closer to the Galactic centre and later migrated outward \citep[e.g.][but see also \citetalias{Dantas2025a}]{Minchev2013, Minchev2018, Frankel2018, Tsujimoto_2020, Lu2024}.

This interpretation is consistent with the inside-out formation scenario of the MW, in which star formation began earlier and proceeded more vigorously in the central regions, producing steep initial metallicity gradients that progressively flattened over time \citep[e.g.][]{Chiappini1997, Magrini2009, Magrini2023}. Consequently, the prevalence of giant-planet hosts with small inferred \rbirth\ values is consistent with both the metallicity dependence of giant-planet formation and the broader picture of the Galaxy’s inside-out formation and growth \citep[a discussion carried out by][]{Haywood_2009}.

In this respect, our findings are also consistent with recent chemo-dynamical models that combine detailed Galactic chemical evolution with radial mixing. \citet[][]{Spitoni2025} suggest that stellar redistribution can enlarge the reservoir of potentially habitable systems across the disc, while the inner Galaxy (owing to its higher metallicity and elevated giant-planet occurrence) may represent a favourable cradle for both massive companions and rocky worlds. The preferential association of giant-planet hosts with inner-disc birth radii provides an observational chemo-dynamical counterpart to this picture.

A similar conclusion was reached by \citet{Boettner2024}, who combined galaxy formation simulations with planet population synthesis models to predict Galactic-scale planet demographics. Their results suggest that the occurrence rate of giant planets is 10--20 times higher in the metal-rich thin disc than in the metal-poor thick disc, whereas low-mass Earth-like planets are comparatively more common in the thick disc. This behaviour arises because giant-planet formation is highly sensitive to the amount of solid material available in the protoplanetary disc, which scales with metallicity, while the formation of small planets seems to depend more weakly on metallicity \citep[][]{Buchhave2012, Mulders2016, Owen2018}.

This distinction is also relevant for Galactic habitability and future technosignature searches. Rocky-only systems in our sample are less centrally concentrated in \rbirth, suggesting that low-mass planets can form across a relatively broad range of Galactic radii. Rocky+giant systems, by contrast, preserve a stronger connection to metal-rich inner-disc birth environments, while also containing rocky planets. Combined with their comparatively older ages, particularly among outward-migrating hosts, these systems provide useful empirical reference populations for identifying long-lived planetary architectures across the Galactic disc (see Table \ref{tab:metallicity_age_deltar}). In this sense, rocky-only and rocky+giant systems may trace complementary search regimes: extended disc environments where rocky planets are common, and inner-disc birth environments where chemically enriched, dynamically redistributed, mixed planetary systems may be found.

\begin{figure}
   \centering
   \includegraphics[width=\linewidth]{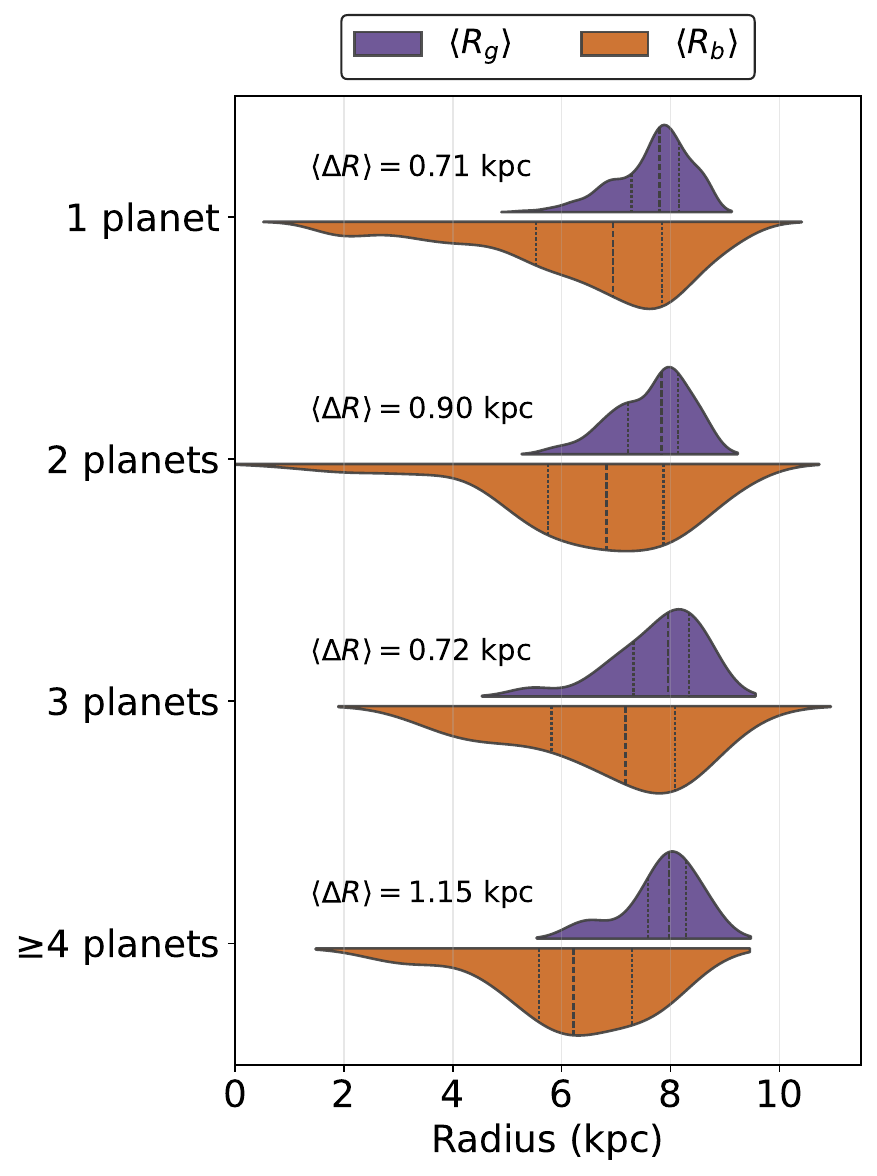}
   \caption{Distributions of guiding $\langle R_g \rangle$ (purple) and birth radii $\langle R_b \rangle$ (orange) for planetary systems grouped by the number of hosted planets, with the median differences $\langle \Delta R \rangle$ annotated for each group, ordered accordingly. The configuration is analogous to Fig. \ref{fig:violin_categories}.}
   \label{fig:violin_num_planets}
\end{figure}

Figure \ref{fig:violin_num_planets} displays asymmetric violin plots for \rgui\ (purple) and \rbirth\ (orange) as a function of the number of hosted planets. The median offset $\langle \Delta R \rangle$ is annotated for each planetary‑class bin. The last bin groups together systems with four or more detected planets, due to their lower individual occurrence. Across all bins we see no clear trend linking the number of planets to either larger or smaller $\langle \Delta R \rangle$ variations.

\subsubsection{How Galactic dynamics may shape planetary systems: I. Insights from orbital actions}
\label{subsubsec:action}

\begin{figure*}
\centering
\includegraphics[width=\linewidth]{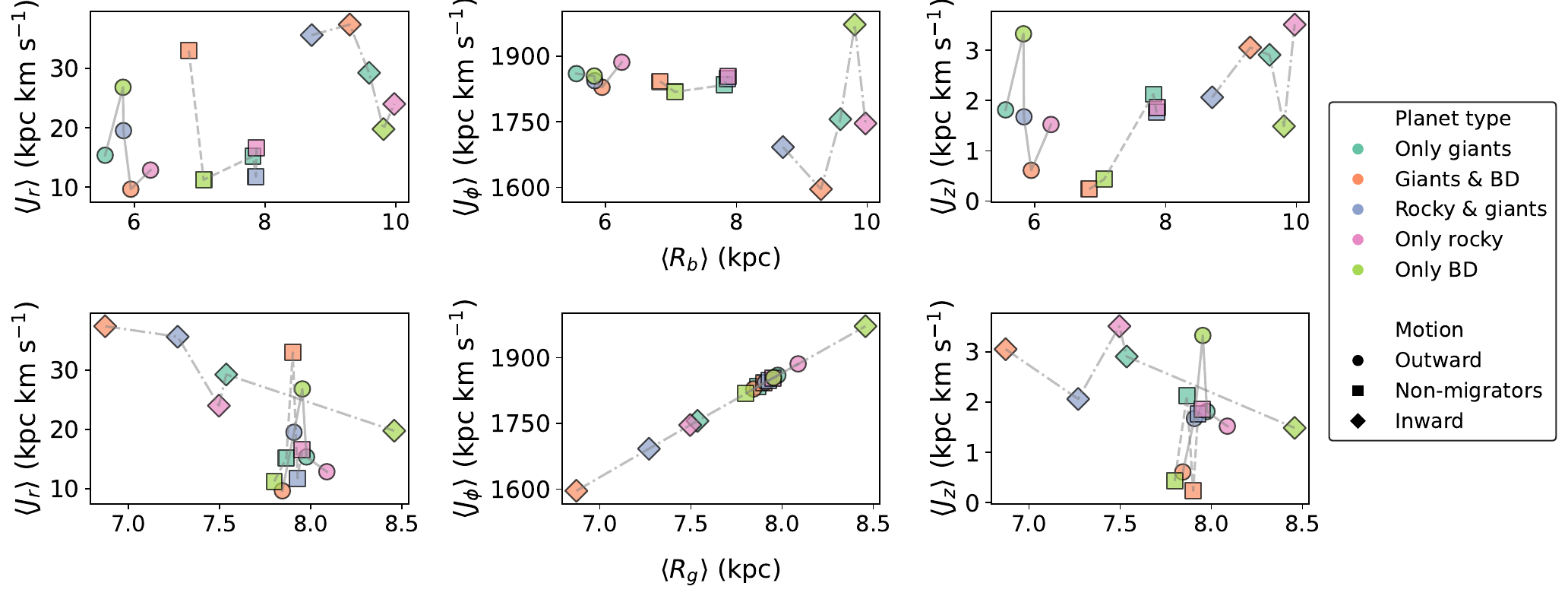}
\caption{Median stellar actions ($\langle J_r \rangle$, $\langle J_\phi \rangle$, $\langle J_z \rangle$) as a function of the median birth radius \rbirth\ (top panels) and guiding radius \rgui\ (bottom panels). Each marker corresponds to the median value of a hosted planet class, colour-coded by planet type. Marker shapes indicate the migration category: outward-migrators, non-migrators, and inward-migrators are represented by circles, squares, and diamonds, respectively. Grey lines connect systems sharing the same motion class.}
\label{fig:actions_vs_radii}
\end{figure*}

The orbital actions (\jr, \jphi, \jz) provide a compact, time-averaged description of stellar orbits: \jr\ traces radial excursions, \jphi\ corresponds to the azimuthal action, essentially the same as the angular momentum about the Galactic symmetry axis, and \jz\ quantifies vertical motion. Figure~\ref{fig:actions_vs_radii} shows the median action components as a function of \rbirth\ and \rgui, separated by planet type and migration class.

The clearest structures are found in \jphi\ and \jz. As expected, \jphi\ follows a one-to-one relation with \rgui, because \rgui\ is set by the angular-momentum content of the orbit. In the \jphi--\rbirth\ plane, the migration classes separate more clearly: outward-migrators tend to combine small \rbirth\ with high \jphi, inward-migrators occupy larger \rbirth\ at lower \jphi\ (although with a much larger variance), and non-migrators lie between these regimes. The separation is less evident in \jz, where all motion classes tend to have larger variance. However, inward-migrators seem to have collectively higher \jz\ than their outward- and non-migrator counterparts.

This behaviour agrees with the enhanced vertical excursions of inward-migrating systems discussed through \zmax\ in Sect.~\ref{subsubsec:additional_diagnostics}. In the present sample, the higher \jz\ values are also consistent with the larger relative contribution of intermediate thin/thick-disc stars among the inward-migrating bins with meaningful statistics, as shown in Table~\ref{tab:metallicity_age_deltar}.

The dependence on planet category is less clear. Some bins involving giant planets or BDs show elevated \jr\ or \jz, but these should not be overinterpreted because several of the relevant category--migration combinations contain few systems. This is particularly important for the inward-migrating Giants \& BDs and only BDs hosts, each represented by a single system in Table \ref{tab:metallicity_age_deltar}. We therefore interpret the action-space structure primarily as a signature of migration history and parent Galactic population, with planet category introducing at most secondary differences. The overall results are in agreement with those of \citetalias[][see Fig. 19 therein]{Dantas2025a}.

\subsubsection{How Galactic dynamics may shape planetary systems: II. Insights from \amax}
\label{subsubsec:amax}

\begin{figure}
   \centering
   \includegraphics[width=\linewidth]{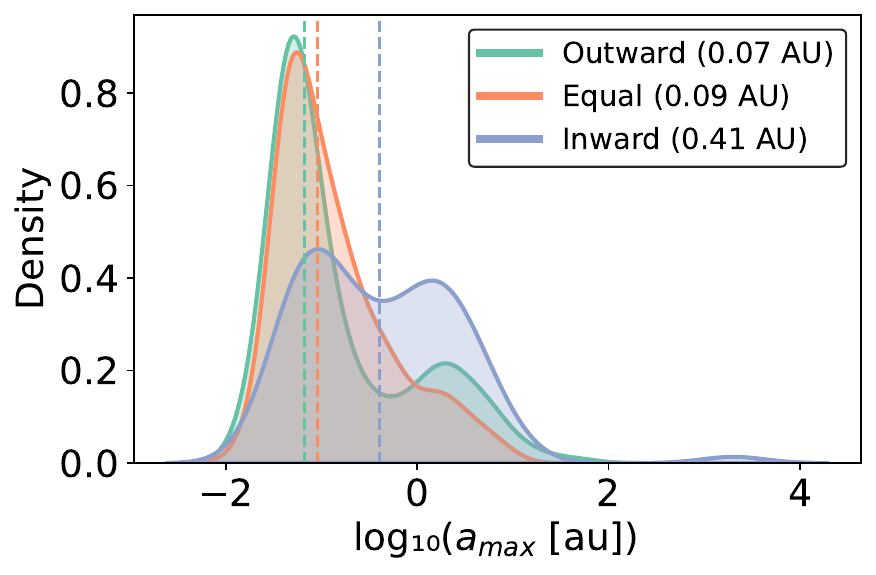}
   \caption{Semi-major axis (\amax) distributions (x-axis in logarithmic scale) for the outermost planet hosted by each star, shown as Gaussian kernel density estimates. Each curve corresponds to one stellar motion class: inward- (blue), non- (orange), and outward-migrators (green). Vertical dashed lines depict the median \amax\ of each subsample, and the legend reports these median values in astronomical units (au).}
   \label{fig:farthest_planets_KDEs}
\end{figure}

Figure \ref{fig:farthest_planets_KDEs} shows Gaussian kernel density estimates of the largest semi-major axis (\amax) among the detected planets in each system, i.e. the orbital distance of the farthest detected companion around a given host star. Outward-migrators display the lowest median \amax\ ($0.07$ au), while inward-migrators exhibit the largest median value ($0.41$ au); non-migrators occupy an intermediate regime ($0.09$ au), although with behaviour closer their outward-moving counterparts.

These differences should be interpreted with caution, because \amax\ is derived from the outermost detected planet in a heterogeneous discovery sample. Transit and radial-velocity detections are intrinsically biased towards short-period, close-in planets owing to geometric probability, signal-to-noise requirements, and finite observing baselines, whereas direct imaging preferentially recovers wide-orbit, massive companions, typically around young stars, and is limited by contrast and angular-separation performance \citep[e.g.][]{Cumming2008, Christiansen2020}. More broadly, the confirmed-planet census aggregates surveys with distinct sensitivities, target selection functions, and follow-up strategies \citep[e.g.][]{Akeson2013}. Therefore, the motion-class differences in median \amax\ may primarily reflect discovery-method demographics and completeness variations rather than an intrinsic dependence of planetary architecture on stellar migration.

If the trend persists after controlling for discovery method and detectability, a dynamical interpretation would become plausible. Wide-orbit planets are more weakly bound and can be destabilised by external perturbations, including stellar encounters and cluster-driven interactions \citep[e.g.][]{Malmberg2011, Hao2013, Parker2020}.  This may be especially relevant for outward migrators, whose smaller \rbirth\ values point to regions of higher stellar density, where such perturbations could be more common. However, the present sample does not support a simple mapping between larger vertical heating and smaller \amax: the most vertically heated hosts are predominantly inward-migrators, whereas the lowest median \amax\ is found among outward-migrators. We therefore refrain from attributing the observed \amax\ sequence to a specific migration-driven heating mechanism. At this stage, the trend should be regarded as a correlation rather than evidence of causation, and as a suggestive architecture--dynamics connection to be tested with discovery-method-controlled subsamples as planet-hosting star samples continue to grow.

Non-migrators provide a useful intermediate comparison, but we avoid assigning their \amax\ distribution to a specific mechanism. Blurring can increase a star's epicyclic amplitude without changing its \rgui\ and may broaden the range of Galactic environments sampled by a planetary system over time \citep[e.g.][]{SellwoodBinney2002, VeraCiro2014, Aumer2016}. Age may further modulate this cumulative exposure \citep[see also][]{Feltzing2020}, but given the heterogeneous detection function and the moderate ages of the non-migrating hosts, we treat blurring and age as possible confounding factors rather than as direct explanations for the observed \amax\ sequence.

\section{Conclusions}
\label{sec:conc}

We analysed 1341 planet-hosting stars, predominantly associated with the Galactic thin disc, to investigate whether Galactic radial migration is reflected in exoplanet-host demographics and planetary-system architectures. Using the birth-radius framework introduced in \citetalias{Dantas2025a}, we compared inferred birth radii (\rbirth) with present-day guiding radii (\rgui) derived from \textsc{Galpy} orbit integrations.

\begin{enumerate}
    \item Most planet-hosting stars in the solar vicinity are consistent with formation at smaller Galactocentric radii than their present-day \rgui, with a smaller fraction inferred to have originated in the outer disc. 

    \item The motion classes follow the expected metallicity sequence: outward-migrators are generally more metal-rich, non-migrators occupy an intermediate regime, and inward-migrators are comparatively more metal-poor. This is consistent with the Galactic radial metallicity gradient and the inside-out growth of the disc.

    \item Giant-planet systems, including mixed giant+BD hosts, are preferentially associated with smaller inferred \rbirth, supporting a link between massive-planet formation and metal-rich inner-disc birth environments. Rocky-only systems show larger characteristic birth radii and smaller collective radial displacements, while BD-only systems appear to occupy a broader and less sharply localised range of birth environments.

    \item Orbital eccentricity does not clearly separate the motion classes: all subsamples remain dominated by low-eccentricity orbits. By contrast, vertical diagnostics reveal dynamically heated planet-hosting stars, including outer-Galaxy-born and intermediate thin--thick disc objects. These high-\zmax\ and high-\jz\ systems likely reflect a combination of age-dependent heating, Galactic-component mixing, radial displacement, and possible external perturbations, rather than radial migration alone.

    \item The farthest detected companion per system shows a tentative dependence on motion class: outward-migrators have the most compact outer detected companions, non-migrators are intermediate, and inward-migrators show the largest median \amax. Although this trend is exploratory, because \amax\ is strongly affected by discovery-method biases and heterogeneous survey completeness, it raises the possibility that outward-migrating systems were originally more extended and subsequently lost or destabilised wide-orbit companions in denser, more dynamically active birth environments. To our knowledge, this is the first explicit indication that Galactic radial migration may be connected to the outer architecture of observed planet-hosting systems.
\end{enumerate}

Our results closely align with those found for the \gaia-ESO field stars in \citetalias{Dantas2025a}, suggesting that planet-hosting stars broadly follow the radial-mixing patterns of the underlying thin-disc population. Additionally, our results indicate that the formation of massive planetary companions is connected to metal-rich inner-Galaxy environments, while the subsequent dynamical evolution of their host stars can redistribute these systems across the disc. The presence of dynamically heated planet-hosting stars further shows that planetary systems can survive substantial Galactic perturbations, although the current data do not yet establish whether their architectures retain a causal imprint of this evolution.

The inferred birth locations and ages of rocky-only and rocky+giant systems are particularly relevant for Galactic habitability. Rocky-only hosts suggest that low-mass planets form over a broader range of Galactic radii, while rocky+giant systems retain a stronger connection to metal-rich inner-disc birth environments. Their comparatively older ages, especially among outward-migrating rocky+giant hosts, make them useful reference populations for future habitability and technosignature searches, as they combine rocky planets, complex architectures, and longer evolutionary timescales.

Future homogeneous samples, with controlled discovery methods, improved stellar ages, and well-characterised completeness functions, will be required to determine whether Galactic dynamics actively reshapes planetary architectures or primarily traces the birth environments in which different planetary systems form.

\section*{Data availability}

The catalogue is only available in electronic form at the CDS via anonymous ftp to \url{cdsarc.u-strasbg.fr} (130.79.128.5) or via \url{http://cdsweb.u-strasbg.fr/cgi-bin/qcat?J/A+A/}.

\begin{acknowledgements}
JJGD acknowledges the European Union through the SPARC-UAH Project under Grant SBPLY/23/180225/000071. MLLD acknowledges Agencia Nacional de Investigación y Desarrollo (ANID), Chile, Fondecyt Postdoctorado Folio 3240344 and ANID Basal Project FB210003. We acknowledge support from ESA through the Science Faculty - Funding reference ESA-SCI-E-LE-281. RS acknowledges support from the National Science Centre, Poland, project 2019/34/E/ST9/00133. This research has made use of the SIMBAD database, operated at CDS, Strasbourg, France. This research has made use of data obtained from or tools provided by the portal \url{https://exoplanet.eu/home/} of the Encyclopaedia of Exoplanetary Systems. This work benefited from the following online platforms: \texttt{slack} (\url{https://slack.com/}), \texttt{github} (\url{https://github.com/}), and \texttt{overleaf} (\url{https://www.overleaf.com/}). Additionally, this work used the following \textsc{python} packages: \textsc{matplotlib} \citep{Hunter2007}, \textsc{numpy} \citep{Harris2020}, \textsc{pandas} \citep{McKinney2010}, \textsc{seaborn} \citep{Waskom2021}, \textsc{mw-plot} (\url{https://milkyway-plot.readthedocs.io/en/stable/}), and \textsc{astropy} \citep{Astropy2013, Astropy2018, Astropy2022}.
\end{acknowledgements}


\bibliographystyle{bibtex/aa}          
\bibliography{bibtex/paper.bib}        

\clearpage

\begin{appendix}
\nolinenumbers  

\section{Supplementary material}
\label{appendix:supplementary}

In this Appendix we present supplementary material that supports and complements the results discussed in the main body of the manuscript.

\subsection{Stellar properties and observational footprint}
\label{appendix:subsec_stellarprop}

\begin{figure}[!ht]
   \centering
   \includegraphics[width=\linewidth]{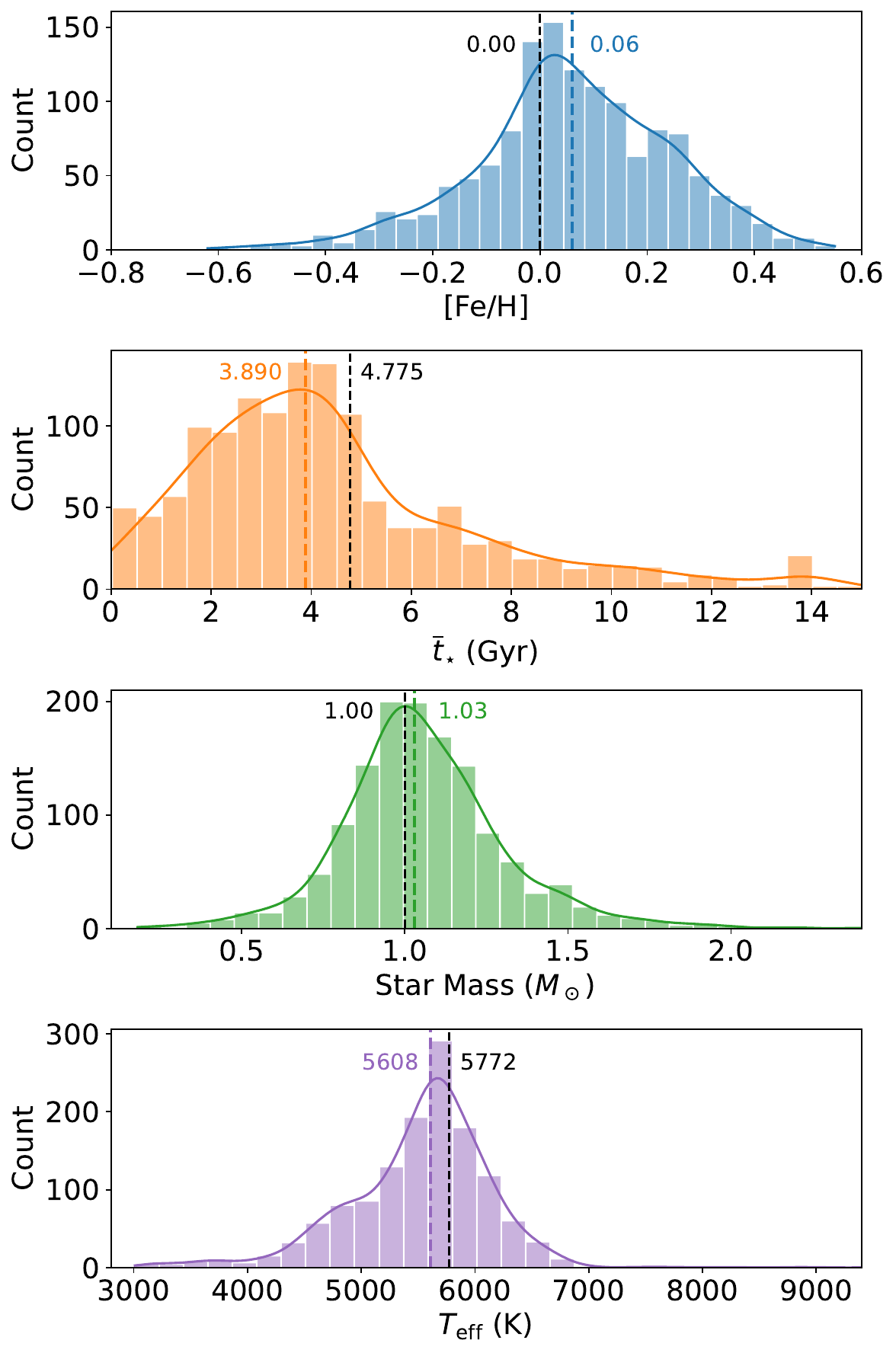}
   \caption{Distributions of key stellar parameters for the final stellar sample. Histograms are shown together with their Gaussian kernel density distributions. Vertical dashed coloured and black dashed lines indicate the median and solar value of each distribution, respectively, with the corresponding values labelled beside them.}
   \label{fig:combined_distributions}
\end{figure}

Figure \ref{fig:combined_distributions} summarises the main stellar properties of the final sample. The host stars are broadly centred around solar metallicity and solar mass, with a median \teff\ close to the solar value. The age distribution peaks at a few Gyr and extends towards older systems, providing the temporal baseline required to investigate long-term Galactic dynamical evolution.

\begin{figure*}[!ht]
   \centering
   \includegraphics[width=\linewidth]{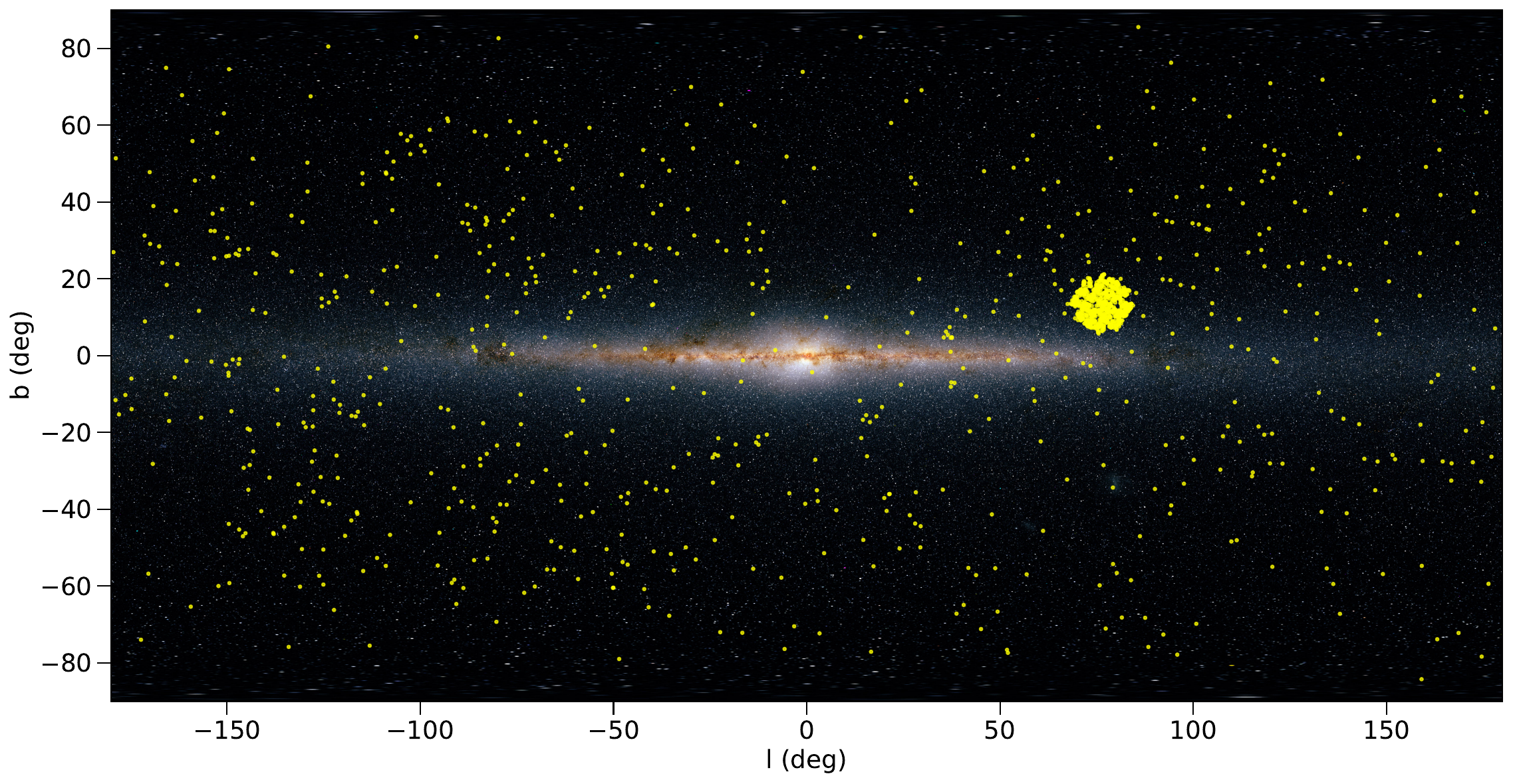}
   \caption{Sky coverage of the final sample of 1341 planet-hosting stars (yellow markers) shown in Galactic coordinates overlaid on the 2MASS near-infrared map (courtesy of ESA/Gaia/DPAC). The prominent overdensity corresponds to the Kepler mission \citep{Borucki2010} footprint.}
   \label{fig:sky_coverage_MW_map}
\end{figure*}

Figure \ref{fig:sky_coverage_MW_map} shows the sky distribution of the sample in Galactic coordinates. Although the stars cover a large fraction of the sky, the prominent overdensity associated with the \textit{Kepler} field illustrates the imprint of major planet-search programmes on the observed exoplanet-host population.

\begin{figure*}[!ht]
    \centering
    \includegraphics[width=\linewidth]{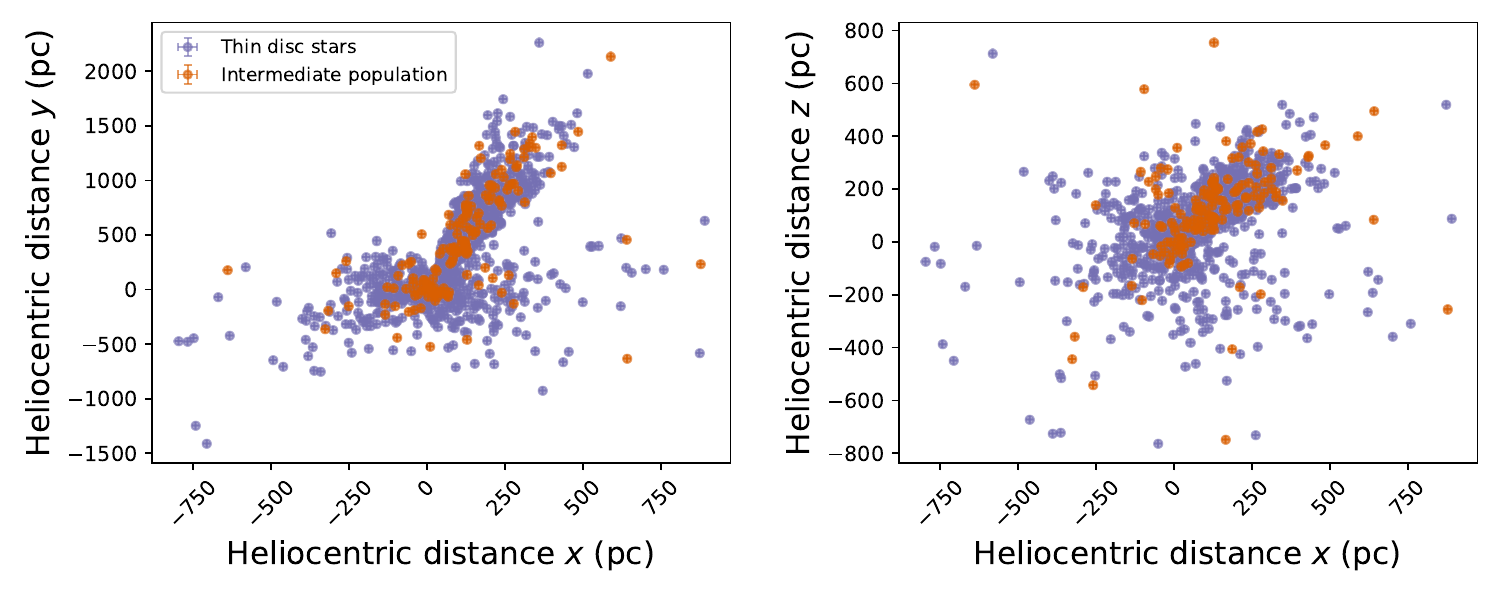}
    \caption{Heliocentric Cartesian distribution of the final planet-hosting sample, shown in the $x$--$y$ plane (left) and $x$--$z$ plane (right). Thin disc stars are shown in purple, while intermediate thin--thick disc objects are shown in orange. The figure illustrates the local spatial footprint of the sample and the overlap between the thin-disc and intermediate populations in the solar neighbourhood.}
\label{fig:heliocentric_distances}
\end{figure*}

Figure \ref{fig:heliocentric_distances} presents the heliocentric Cartesian distances of the final sample. The stars occupy a local volume around the Sun, with thin-disc and intermediate-population objects partially overlapping in projected space.

\subsection{Radial and vertical dynamical diagnostics}
\label{appendix:subsec_dynamical_diagnostics}

\begin{figure}[!ht]
   \centering
   \includegraphics[width=\linewidth]{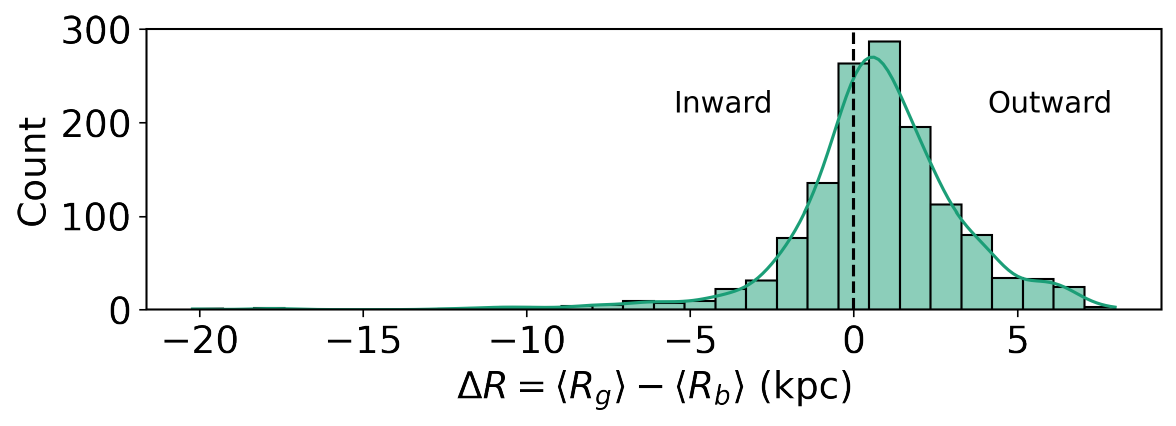}
   \caption{Distribution (histogram and Gaussian kernel density) of \mbox{$\Delta R \equiv $ \rgui\ $-$ \rbirth} for the final sample (excluding thick disc stars). The vertical dashed black line marks $\Delta R=0$. Labels on either side of the line indicate the direction of inward or outward motion.}
   \label{fig:migration_distr}
\end{figure}

Figure \ref{fig:migration_distr} shows the distribution of radial displacement, $\Delta R \equiv$ \rgui $-$ \rbirth, for the final sample after excluding thick-disc stars. The distribution is concentrated around small offsets but includes extended tails, indicating that the solar-vicinity planet-host sample contains both systems with little net radial displacement and systems whose inferred birth radii differ substantially from their present-day guiding radii.

\begin{figure*}[!ht]
    \centering
    \includegraphics[width=\linewidth]{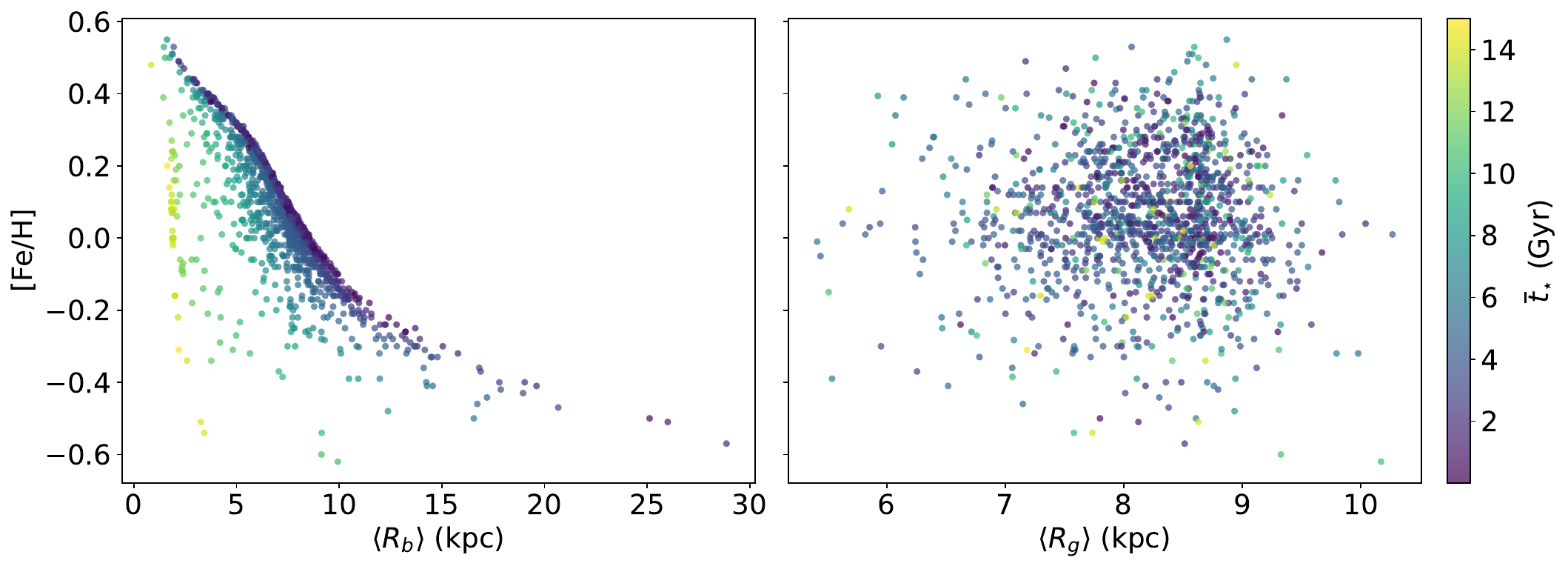}
    \caption{Birth radius (\rbirth) and guiding radius (\rgui) as a function of \feh\ for the planet-hosting stars, colour-coded by stellar age (\age). Left panel: estimated birth radii, showing clear metallicity and age gradients. Right panel: present-day guiding radii, illustrating how radial mixing reduces the correlation between metallicity, age, and Galactocentric radius.}
    \label{fig:rb_rg_metal_ages}
\end{figure*}

Figure \ref{fig:rb_rg_metal_ages} compares \rbirth\ and \rgui\ as a function of \feh, colour-coded by stellar age. The \rbirth--\feh\ plane shows the expected imprint of the Galactic radial metallicity gradient: metal-rich stars are preferentially associated with smaller inferred birth radii, while metal-poor stars extend towards larger \rbirth. In contrast, the \rgui--\feh\ plane is more mixed, illustrating how radial redistribution weakens the present-day relation between metallicity and Galactocentric radius.

\begin{figure*}[!ht]
    \centering
    \includegraphics[width=\linewidth, trim={3.5cm 0 1.cm 0}, clip]{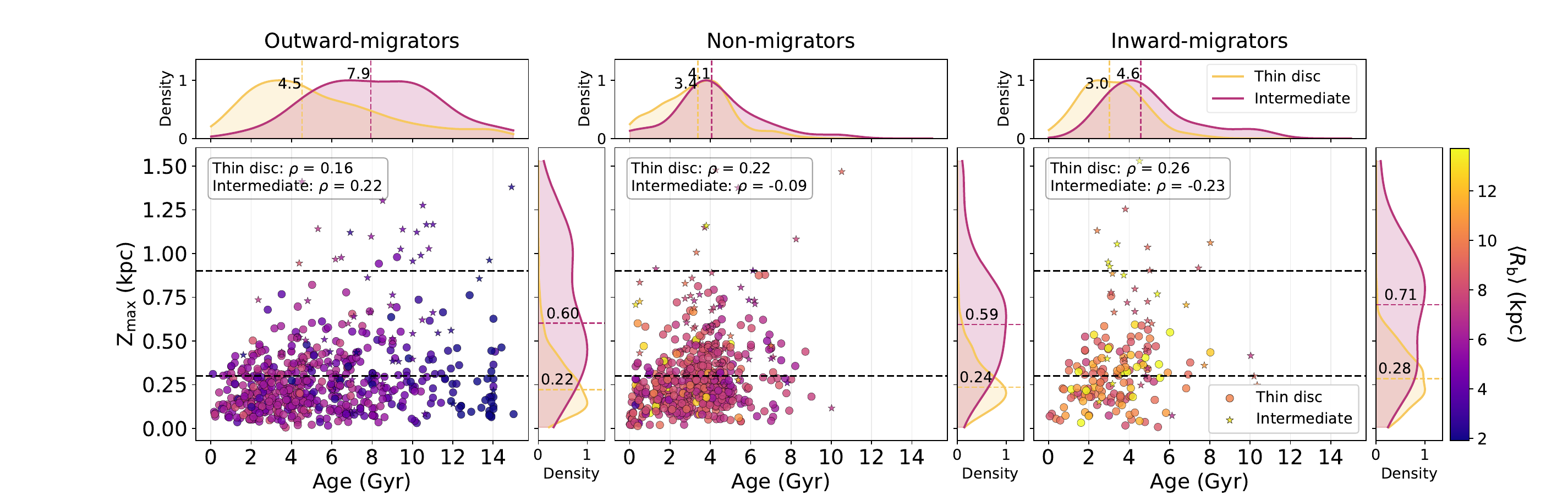}
    \caption{Median maximum Galactic scale height (\zmax) as a function of stellar age (\age), split by radial-motion class. From left to right, the panels show outward-, non-, and inward-migrators. Points are colour-coded by \rbirth, while marker shape distinguishes thin disc stars from intermediate thin--thick disc objects. Marginal kernel density estimates show the age and \zmax\ distributions for the thin-disc and intermediate populations separately. Horizontal dashed lines mark the canonical thin-disc and thick-disc scale-height thresholds at $Z_{\rm max}=300$ and 900 pc, respectively \citep{McMillan2017}. The Spearman rank coefficient reported in each panel quantifies the age--\zmax\ correlation within each Galactic component.}
    \label{fig:zmax_age_components}
\end{figure*}

Figure \ref{fig:zmax_age_components} examines the relation between \zmax\ and \age\ across the three radial-motion classes. Most systems remain within thin-disc-like vertical excursions, but a subset reaches larger \zmax, including several intermediate-population stars. The age--\zmax\ trends differ between the thin-disc and intermediate components, supporting the interpretation that the vertical signal likely reflects a combination of dynamical heating, component mixing, and potential interactions with other systems (e.g. Sagittarius), rather than a single mechanism.

\subsection{Features of the exoplanet population in our sample}
\label{appendix:subsec:exoplanet_features}

This subsection summarises key statistical properties of the exoplanet host stars in relation to their inferred \rbirth, present-day \rgui, \feh, \age, and dynamical classifications. These complementary diagnostics provide a compact overview of how different planetary properties map onto Galactic structure and stellar migration pathways.

\begin{figure*}[!ht]
    \centering
    \includegraphics[width=\linewidth]{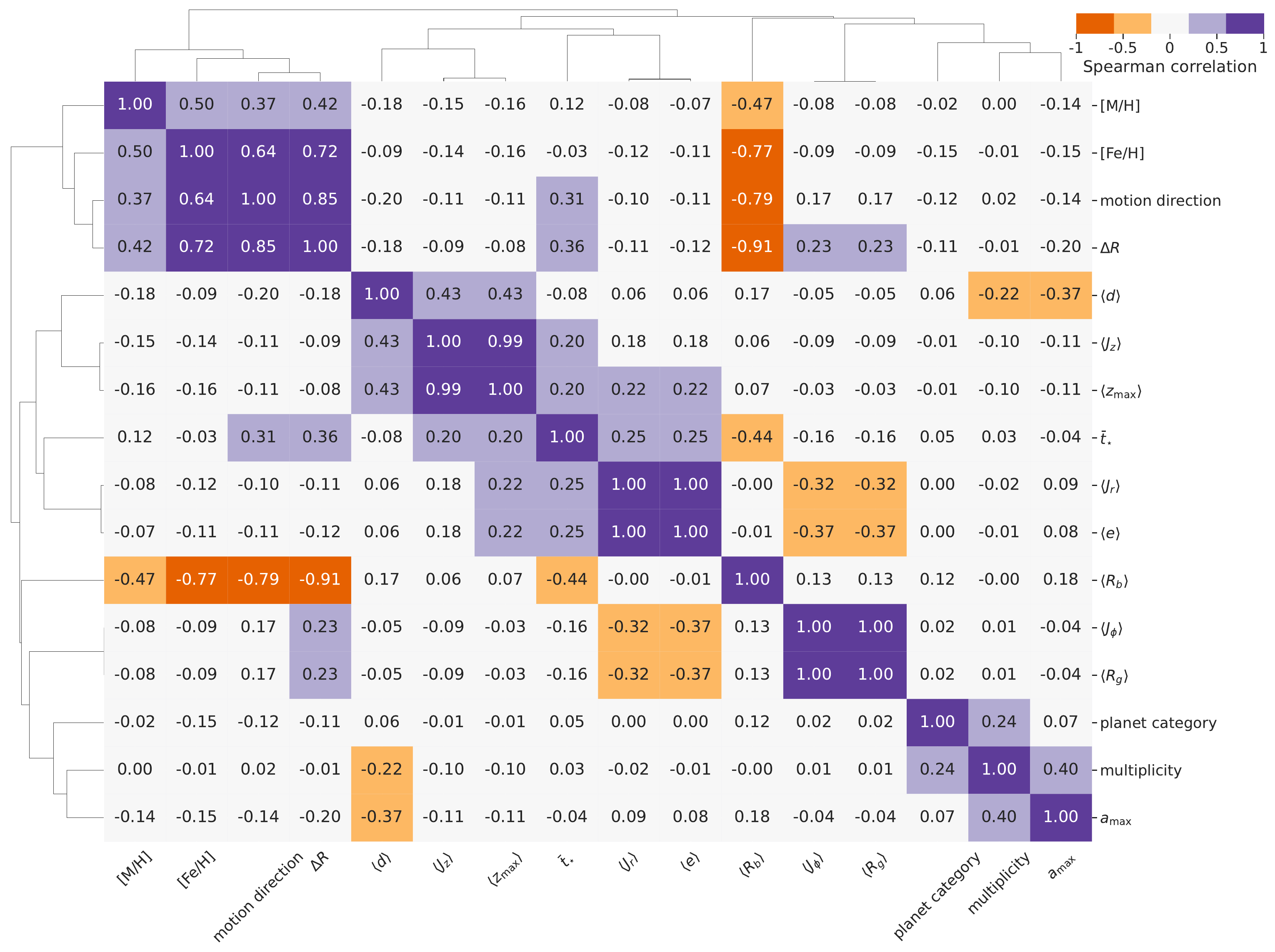}
    \caption{Clustermap showcasing the Spearman rank correlation coefficients between selected stellar, orbital, and planetary parameters in the sample. The matrix is hierarchically clustered to highlight groups of quantities with similar monotonic behaviour. The strongest correlations mostly trace expected relationships among \rbirth, \rgui, $\Delta R$, metallicity, motion class, and angular-momentum-related orbital quantities. Correlations involving encoded categorical parameters are descriptive and depend on the adopted ordering.}
    \label{fig:heatmap}
\end{figure*}

Table \ref{tab:stats_rb_rg} reports descriptive statistics of \rbirth\ and \rgui\ across planetary categories, allowing a direct comparison of their radial distributions. Table \ref{tab:metallicity_age_deltar} further separates each category by direction of radial motion, linking chemical enrichment and \age\ to motion history. Finally, Fig. \ref{fig:heatmap} presents the rank correlations between stellar and planetary parameters, offering a global view of the interdependencies within the sample.

Figure \ref{fig:heatmap} summarises the main monotonic correlations among the stellar, orbital, and planetary quantities used throughout the analysis. The strongest correlations are associated with the construction of the radial-migration diagnostics themselves: \rbirth, \rgui, $\Delta R$, direction of motion, and metallicity are tightly linked through the Galactic radial metallicity gradient and the definition $\Delta R \equiv$ \rgui $-$ \rbirth. \jz\ and \rgui\ also cluster together, as expected from their common dependence on the angular-momentum content of the orbit. By contrast, planetary multiplicity shows no meaningful monotonic relation with $\Delta R$, consistent with the absence of a clear trend between radial displacement and the number of detected planets. Correlations involving encoded categorical variables, such as planetary category, Galactic component, or direction of motion, should be interpreted only as descriptive summaries of the adopted ordering, rather than as direct physical phenomena.

\begin{table*}[!ht]
    \caption{Descriptive statistics of the inferred birth radius (\rbirth) and present-day guiding radius (\rgui) for each planetary category. Reported quantities include the number of systems, mean ($\mu$), standard deviation ($\sigma$), selected percentiles, extrema, and median absolute deviation (MAD). Radii are given in kpc.}
    \centering
    \label{tab:stats_rb_rg}
    \begin{tabular}{lcrrrrrrrrrrrrrr}
    \toprule
    Planet category & Radius & \# & $\mu$ & $\sigma$ & min & .15\% & 2.5\% & 16\% & 50\% & 84\% & 97.5\% & 99.85\% & max & MAD \\
    \midrule
    \multirow[c]{2}{*}{Giants \& BD} 
     & \rbirth & \multirow[c]{2}{*}{16} & 6.3 & 1.4 & 2.3 & 2.4 & 3.5 & 5.5 & 6.3 & 7.3 & 8.6 & 9.3 & 9.3 & 0.6 \\
     & \rgui   &  & 7.6 & 0.7 & 6.3 & 6.3 & 6.4 & 6.9 & 7.8 & 8.1 & 8.5 & 8.6 & 8.6 & 0.2 \\    
    \midrule
    \multirow[c]{2}{*}{Only BD} 
     & \rbirth & \multirow[c]{2}{*}{13} & 6.2 & 1.9 & 2.9 & 2.9 & 3.1 & 4.2 & 6.5 & 7.8 & 9.3 & 9.8 & 9.8 & 1.2 \\
     & \rgui   & & 7.9 & 0.8 & 6.6 & 6.6 & 6.6 & 6.9 & 8.0 & 8.6 & 9.3 & 9.4 & 9.4 & 0.5 \\     
    \midrule
    \multirow[c]{2}{*}{Only giants} 
     & \rbirth & \multirow[c]{2}{*}{966} & 7.0 & 2.7 & 0.8 & 1.5 & 2.0 & 4.8 & 7.0 & 8.6 & 12.8 & 25.6 & 28.9 & 1.3 \\
     & \rgui   &  & 7.8 & 0.8 & 4.9 & 5.3 & 6.2 & 7.0 & 7.9 & 8.6 & 9.1 & 10.0 & 10.5 & 0.5 \\     
    \midrule
    \multirow[c]{2}{*}{Rocky \& giants} 
     & \rbirth & \multirow[c]{2}{*}{93} & 7.1 & 2.1 & 1.9 & 2.1 & 3.4 & 5.3 & 7.3 & 8.3 & 10.9 & 17.2 & 17.9 & 1.0 \\
     & \rgui   &  & 7.7 & 0.7 & 5.3 & 5.4 & 6.1 & 7.0 & 7.9 & 8.4 & 8.8 & 9.3 & 9.4 & 0.5 \\     
    \midrule
    \multirow[c]{2}{*}{Only rocky} 
     & \rbirth & \multirow[c]{2}{*}{253} & 7.7 & 2.6 & 1.7 & 1.8 & 1.9 & 6.1 & 7.8 & 9.4 & 14.2 & 19.2 & 20.7 & 0.9 \\
     & \rgui   &  & 7.9 & 0.8 & 5.2 & 5.4 & 6.2 & 7.0 & 8.0 & 8.6 & 9.6 & 10.2 & 10.3 & 0.5 \\
    \bottomrule
    \end{tabular}
\end{table*}


\begin{table*}[!ht]
    \caption{Median metallicity ($\langle \rm{\feh} \rangle$), median stellar age (\age), and radial-displacement diagnostics for each planetary category, subdivided by direction of radial motion (inward, outward, and no net migration). The quantities $W_{68}(\Delta R)$ and $W_{90}(\Delta R)$ correspond to the central 68 and 90 per cent widths of the $\Delta R$ distribution, respectively.}
    \centering
    \begin{tabular}{ll S[table-format=3.0] S[table-format=3.0] S[table-format=1.2] S[table-format=2.2] S[table-format=2.2] S[table-format=2.2] S[table-format=2.2] S[table-format=2.2]}
    \toprule
    Hosted planet category & Direction of motion & {$N_{\rm{thin}}$} & {$N_{\rm{intermediate}}$} & {$\langle \rm{\feh} \rangle$} & {$\overline{t}_{\star}$} & {$\Delta R_{\rm min}$} & {$\Delta R_{\rm max}$} & {$W_{68}(\Delta R)$} & {$W_{90}(\Delta R)$} \\
    & & & & {dex} & {Gyr} & {kpc} & {kpc} & {kpc} & {kpc} \\
    \midrule
    \multirow{3}{*}{Giants \& BDs} & Outward          & 12 & 0 &  0.22 & 4.40 & -5.60 & -0.21 & 1.56 & 3.34 \\
                                   & No net migration &  3 & 0 &  0.14 & 1.45 & -1.07 &  0.18 & 0.85 & 1.12 \\
                                   & Inward           &  1 & 0 & -0.28 & 6.90 &  2.42 &  2.42 &  N/A & N/A \\
    \midrule
    \multirow{3}{*}{Only BDs} & Outward          & 5 & 2 &  0.14 & 4.60 & -4.78 & -0.64 & 3.57 & 3.97 \\
                              & No net migration & 5 & 0 &  0.13 & 2.10 & -2.75 &  0.41 & 1.86 & 2.75 \\
                              & Inward           & 1 & 0 & -0.09 & 0.69 &  1.35 &  1.35 &  N/A &  N/A \\
    \midrule
    \multirow{3}{*}{Only giants} & Outward          & 417 & 44 &  0.19 & 4.79 & -7.99 & -0.32 & 3.05 & 5.24 \\
                                 & No net migration & 361 & 39 &  0.02 & 3.26 & -4.62 & 20.21 & 1.60 & 3.83 \\
                                 & Inward           &  86 & 19 & -0.15 & 3.24 &  0.35 & 18.03 & 3.02 & 8.15 \\
    \midrule
    \multirow{3}{*}{Rocky \& giants} & Outward          & 36 & 6 &  0.17 & 6.70 & -6.82 & -0.60 & 1.79 & 4.06 \\
                                     & No net migration & 34 & 2 &  0.00 & 3.64 & -1.58 &  1.87 & 1.24 & 2.25 \\
                                     & Inward           & 11 & 4 & -0.14 & 3.80 &  0.33 &  9.52 & 2.27 & 6.14 \\
    \midrule
    \multirow{3}{*}{Only rocky} & Outward          &  71 &  7 &  0.13 & 4.57 & -6.79 & -0.68 & 4.10 & 5.37 \\
                                & No net migration & 113 & 16 &  0.02 & 3.89 & -3.15 & 12.38 & 1.66 & 3.05 \\
                                & Inward           &  34 & 12 & -0.16 & 3.32 &  0.93 & 10.59 & 4.60 & 6.18 \\
    \bottomrule
    \end{tabular}
    \label{tab:metallicity_age_deltar}
\end{table*}

\end{appendix}

\end{document}